\documentclass[11pt]{article}
\usepackage{amsmath,amssymb}
\usepackage{amsthm}
\usepackage{ascmac}
\usepackage{float}
\usepackage{bm}
\usepackage[dvipdfmx]{graphicx}
\usepackage{comment}
\usepackage[all]{xy}
\usepackage[bookmarksnumbered=true,colorlinks=true,filecolor=blue,linkcolor=blue,urlcolor=blue,citecolor=red,linktocpage=true]{hyperref}

\allowdisplaybreaks

\numberwithin{equation}{section}

\newcommand{\re}{\mathrm{e}}
\newcommand{\ri}{\mathrm{i}}

\newcommand{\be}{\begin{equation}}
\newcommand{\ee}{\end{equation}}
\newcommand{\ba}{\begin{aligned}}
\newcommand{\ea}{\end{aligned}}

\usepackage{ulem}

\begin{document}


\renewcommand{\thefootnote}{\fnsymbol{footnote}}
\setcounter{page}{0}
\thispagestyle{empty}
\begin{flushright}  OCU-PHYS 487 \end{flushright} 

\vskip3cm
\begin{center}
{\LARGE {\bf Quantum mirror curve of periodic chain geometry}}
 \vskip1.5cm
{\large 
 {
 \sc Taro Kimura%
 \footnote{\href{mailto:taro.kimura@keio.jp}{\tt taro.kimura@keio.jp}}%
and
 \sc Yuji Sugimoto%
 \footnote{\href{mailto:sugimoto@sci.osaka-cu.ac.jp}{\tt sugimoto@sci.osaka-cu.ac.jp}}}

\vskip1.5cm
\it $^{*}$Department of Physics, Keio University, Kanagawa 223-8521, Japan

\vskip.5cm
\it $^{\dagger}$Osaka City University Advanced Mathematical Institute (OCAMI), \\ 3-3-138, Sugimoto, Sumiyoshi-ku, Osaka, 558-8585, Japan} 
\end{center}

\vskip1cm
\begin{abstract} 
 The mirror curves enable us to study B-model topological strings on non-compact toric Calabi--Yau threefolds. One of the method to obtain the mirror curves is to calculate the partition function of the topological string with a single brane. In this paper, we discuss two types of geometries: one is the chain of $N$ $\mathbb{P}^1$'s which we call ``$N$-chain geometry,'' the other is the chain of $N$ $\mathbb{P}^1$'s with a compactification which we call ``periodic $N$-chain geometry.'' We calculate the partition functions of the open topological strings on these geometries, and obtain the mirror curves and their quantization, which is characterized by (elliptic) hypergeometric difference operator.
 We also find a relation between the periodic chain and $\infty$-chain geometries, which implies a possible connection between 5d and 6d gauge theories in the larte $N$ limit.
\end{abstract}

\renewcommand{\thefootnote}{\arabic{footnote}}
\setcounter{footnote}{0}

\vfill\eject

\tableofcontents

\section{Introduction and Summary}\label{Intro}
Topological string introduced in \cite{Witten} is known as a toy model to understand string theory compactified on Calabi--Yau threefolds.
Two formulations, the A-model topological string which depends on the K\"ahler moduli, and the B-model topological string depending on the complex modulus, are related to each other via mirror symmetry \cite{Candelas}.
For the non-compact toric Calabi--Yau threefolds, the gravity sector is decoupled, and it ends up with gauge theory with supersymmetry:
The A-model topological string computes the partition function of 5d $\mathcal{N}=1$ supersymmetric gauge theories compactified on a circle with the self-dual omega background ~\cite{Katz:1996fh}. In the B-model topological strings, the partition function is closely related to the entropy of the supersymmetric black hole ~\cite{Ooguri:2004zv}. The B-model topological sring is encoded to the complex 1-dimensional manifold in the toric Calabi--Yau threefold $\Sigma = \{ (x,p) \in \mathbb{C}^* \times \mathbb{C}^* \mid H(x,p)=0 \}$ where $x,p \in \mathbb{C}^* = \mathbb{C}\backslash\{0\}$, the so-called ``mirror curve''~\cite{Aganagic:2003qj}.

Recently, it has been shown that quantization of the mirror curves introduced in \cite{Aganagic:2003qj, Aganagic:2011mi} provides the non-perturbative effects in the B-model topological strings on the toric Calabi--Yau threefolds~\cite{Grassi:2014zfa}. 
This formalism is now known to be relevant to ABJM theory \cite{Moriyama:2014gxa, Moriyama:2017nbw}, integrable system \cite{Hatsuda:2015qzx, Franco:2015rnr}, and even condensed matter physics \cite{Hatsuda:2016mdw, Hatsuda:2017zwn}. Therefore, it seems worthy to study the mirror curve and its quantization for various fields.

\begin{figure}[htb]
\begin{minipage}{0.5\hsize}
\includegraphics[width=8cm]{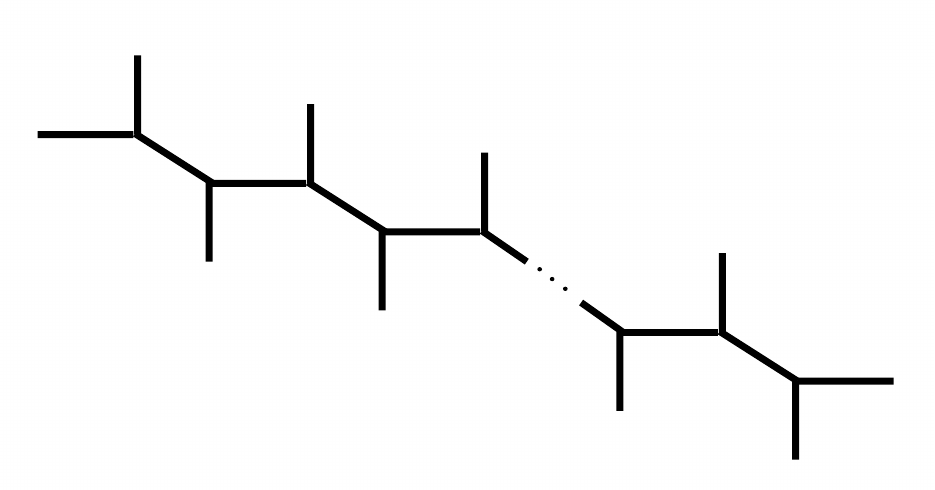}
\end{minipage}
\put(-225,65){(a)}
\hspace{8mm}
\begin{minipage}{0.45\hsize}
\includegraphics[width=8cm]{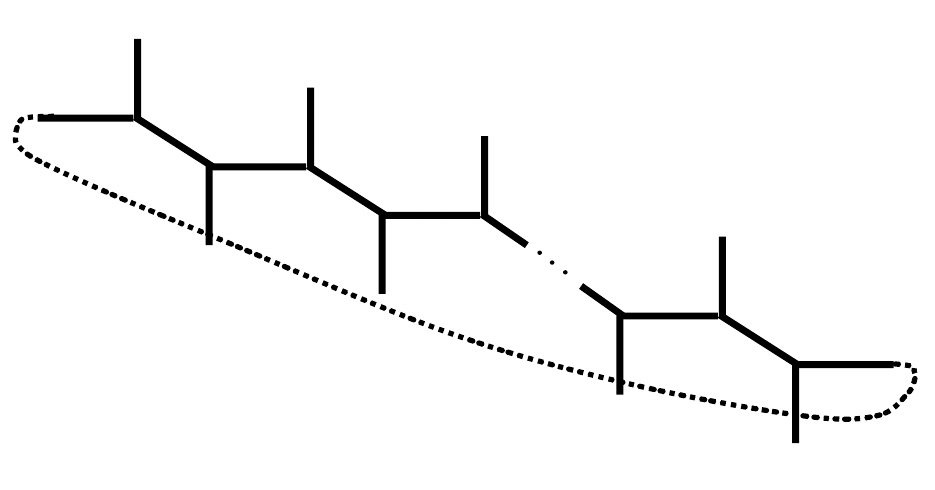}
\end{minipage}
\put(-205,65){(b)}
\caption{The web diagram descriptions of the toric Calabi--Yau threefolds. The left figure (a) is the $N$-chain geometry which contains $N$ vertical lines. The right figure (b) is the periodic $N$-chain geometry. The black dotted lines imply the compactification.}
\label{CYs}
\end{figure}

In this paper, 
we consider the topological string theories on two kinds of toric Calabi--Yau threefolds:
One is the chain geometry which is defined as a hirizontal chain of $\mathbb{P}^1$'s that we call ``$N$-chain geometry.''
The other is the chain of $\mathbb{P}^1$'s as well, but the horizontal axis is compactified, that we call ``periodic $N$-chain geometry'', as in Fig.~\ref{CYs}\footnote{A different type of periodic geometry is considered in, e.g. ~\cite{Kim:2017jqn}.}. The left panel of Fig.~\ref{CYs} is studied, e.g. in~\cite{Iqbal:2004ne, Hwang:2012jh}. The quantized mirror curves are discussed, e.g. in ~\cite{Bonelli:2011fq, Gukov:2011qp, Halmagyi:2005vk}.  The right panel of Fig.~\ref{CYs} is studied in the context of 6 dimensional gauge theories~\cite{Hollowood:2003cv, Sugimoto:2015nha, Haghighat:2013gba, Haghighat:2013tka}, and the Seiberg--Witten curve for the similar case has been studied in~\cite{Haghighat:2016jjf}.

The partition function of the open topological string $\mathcal{Z}^{\text{o}}(x)$ behaves as the wave function which is annnihilated by an operator $\mathsf{H}(\mathsf{x},\mathsf{p})$,
\be
\ba
\mathsf{H}(\mathsf{x},\mathsf{p}) \mathcal{Z}^{\text{o}}(x)=0, \qquad
[\log \mathsf{p}, \log \mathsf{x}] = \ri g_s
\ea
\ee
where $x$ is the modulus of brane, and $g_s$ is the string coupling constant \cite{Aganagic:2003qj, Aganagic:2011mi}. We call above difference equation as ``quantum mirror curve.'' The operator $\mathsf{H}(\mathsf{x},\mathsf{p})$ can be interpreted as the quantization of the polynomial $H(x,p)$. Conversely, by replacing the operators $\mathsf{x}$ and $\mathsf{p}$ with the classical variables $x$ and $p$, we obtain a polynomial $H(x,p)$ and the classical curve $\Sigma = \{ (x,p) \mid H(x,p)=0 \}$\footnote{In general, due to the ambiguity of the order of the operators $\mathsf{x}$ and $\mathsf{p}$, $H(\mathsf{x},\mathsf{p})$ is not the same as $\mathsf{H}(\mathsf{x},\mathsf{p})$.}. We can calculate $\mathcal{Z}^{\text{o}}(x)$ by the topological vertex technique \cite{Aganagic:2003db, Iqbal:2007ii}\footnote{In this paper we consider the unrefined topological strings. The brane insertion in the refined topological strings is discussed in \cite{Mori:2016qof, Kimura:2017auj, Kameyama:2017ryw}. The non-perturbative effects of the open topological string was calculated in \cite{Sugimoto:2016vnb}.}, and derive the mirror curve in both cases in Section~\ref{mirror_curve}, which is consistent with the known result especially for $N =1$.

In this paper we calculate the partition functions of the topological string with the brane on vertical axis and horizontal axis, $\mathcal{Z}^{\text{o}}(x)$ and ${\mathcal{Z}'}^{\text{o}}(x)$. Accordingly, we derive two operators annihilating the partition functions, $\mathsf{H}(\mathsf{x},\mathsf{p})$ and $\mathsf{H}'(\mathsf{x},\mathsf{p})$, and take the classical limit. Then we find
\be
\ba
\mathsf{H}(q^{-1/2} Q_m^{-1}  \mathsf{p},q^{1/2}\mathsf{x})\Big|_{Q_m\to  Q_m^{-1} q^{-1}}=\mathsf{H}'(\mathsf{x}, \mathsf{p}), \qquad
H(Q_m^{-1} p,x)\Big|_{Q_m \to Q_m^{-1}}={H'}(x,p).
\vspace{10mm}
\ea
\ee
under some constraints for the K\"ahler parameters denoted by $Q_m$ where $q:=\re^{\ri g_s}$.

It has been shown in \cite{Hollowood:2003cv} that the toric diagram of periodic chain geometry is given by a set of the infinite chain geometry, as we will also mention in Section~\ref{mirror_curve}. Based on this result, we expect the partition functions for the chain geometry to be related with the one for the periodic conifold. Then, we find that the partition function of the \textit{open} topological string on the periodic chain geometry is exactly the same as the one on the $\infty$-chain geometry with the brane inserted on the internal line. This means that the mirror curve of the periodic chain geometry can be realized as the $\infty$-chain geometry. In the context of the supersymmetric gauge theories geometrically engineerd from the toric Calabi--Yau threefolds~\cite{Iqbal:2003zz, Eguchi:2003sj, Dimofte:2010tz, Taki:2010bj, Awata:2010bz, Bonelli:2011fq}, the compactified theories with the defect can be obtained from the uncompactified theories with the defect. We stress that although the partition function of the \textit{closed} topological string on periodic chain geometry does not agree with the one on $\infty$-chain geometry, we find a relation between them.

The rest of the paper is organized as follows. In Section~\ref{calcPF}, we calculate the partition functions of the topological strings with the branes on (periodic) $N$-chain geometries by utilizing the topological vertex. Actual computation is based on the vertex operator formalism which is well established in the context of the melting crystal~\cite{Okounkov:2003sp, Okuda:2004mb}. In Section \ref{mirror_curve}, we will provide the prescription to obtain the mirror curves and its toric diagrams corresponding to the web diagrams by the method argued in~\cite{Taki:2010bj,Takasaki:2015raa}. In Section~\ref{discussion}, we conclude with some remarks and discussions. The notation and formula are summarized in Appendix~\ref{appA}.

\section{Chain geometries and its compactification}\label{calcPF}

In this Section we calculate the topological string partition function with additional branes on two kinds of the toric Calabi--Yau threefolds, that we call the chain geometries, based on the topological vertex formalism.
In practice, we apply the vertex operator formalism to compute the partition function.

\subsection{Chain geometries}

\begin{figure}[htb]
\begin{minipage}{0.5\hsize}
\includegraphics[width=8cm]{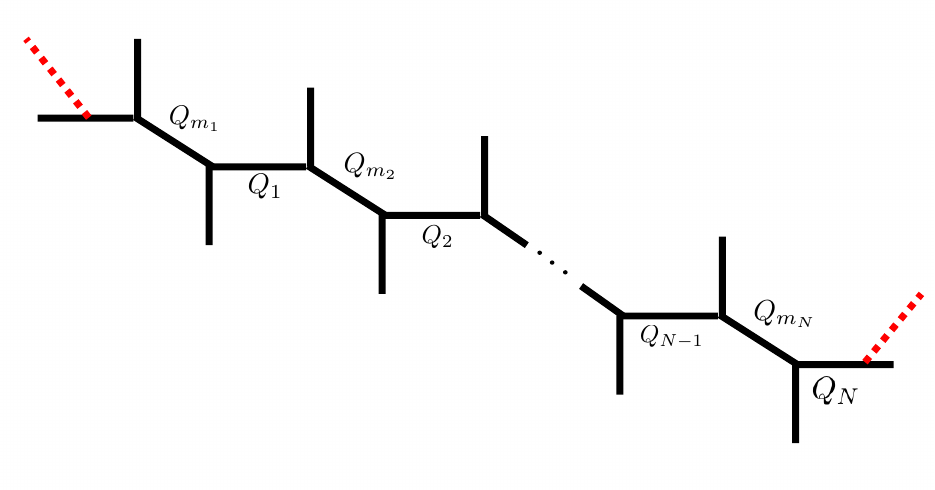}
\end{minipage}
\put(-225,65){(a)}
\hspace{8mm}
\begin{minipage}{0.45\hsize}
\includegraphics[width=8cm]{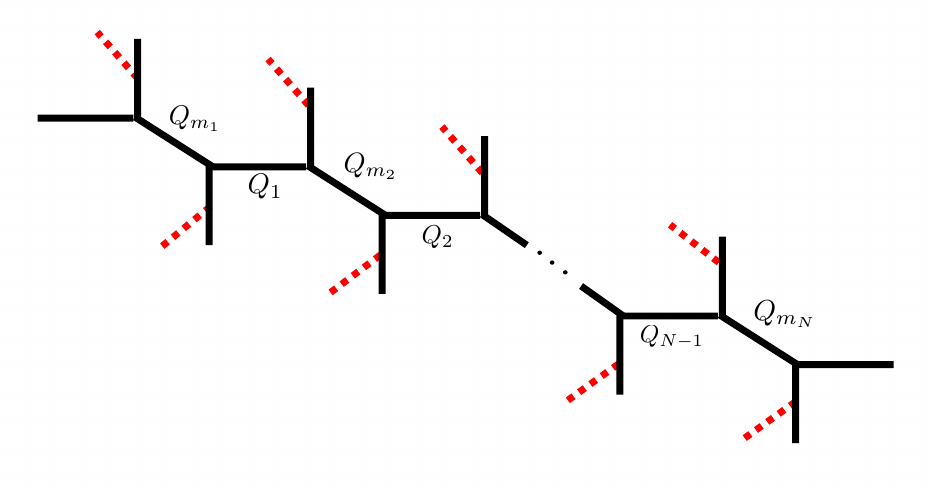}
\end{minipage}
\put(-205,65){(b)}
\caption{The $N$-chain geometries with the branes denoted by the red lines. The variables $Q_{m_i}$ and $Q_i$ denote the K\"ahler parameters along the slanting and horizontal lines.}
\label{chain}
\end{figure}

We consider the $N$-chain geometry with additional branes on the left and right most of horizontal lines as shown in Fig.~\ref{chain} (a).
The corresponding topological string partition function is given by

\be
\ba
\mathcal{Z}_{\text{chain}}^{N}
=
\sum_{\text{all indices}}
{\rm Tr}_{\mu_0^t} U
\prod_{a=1}^N \Bigl[ 
(-Q_{m_a})^{|\mu_{m_a}|} (-Q_{a} )^{|\mu_a|}
C_{\mu_{a-1}^t \mu_{m_a} \emptyset}  C_{\mu_a \mu^t_{m_{a}} \emptyset}  \Bigr]
{\rm Tr}_{\mu_N}W,
\label{pf_chain}
\ea
\ee
where $U$ and $W$ are the holonomy matrices associated with the branes. The function $C_{\lambda \mu \nu}(q)$ is the topological vertex defined as 
\be
\ba
&C_{\lambda \mu \nu}(q) := q^{\frac{\kappa_{\mu}+\kappa_{\nu}}{2}}s_{\nu}(q^{-\rho})
\sum_{\eta} s_{\lambda^{t}/\eta}(q^{-\rho-\nu})s_{\mu/\eta}(q^{-\rho-\nu^{t}}),
\label{vertex}
\ea
\ee
where $s_{\mu/\eta}(x)$ is the skew-Schur function, and $\rho$ is the Weyl vector. The variable, $\kappa_\mu$ is defined in Appendix \ref{appA}. By substituting \eqref{vertex} to \eqref{pf_chain} and using ${\rm Tr}_\mu M = s_\mu (z)$ where $z=\{z_i\}$ is the set of the eigenvalues of the holonomy matrix $M$, we find
\be
\ba
&\mathcal{Z}_{\text{chain}}^N
\\
&=
\sum_{\text{all indices}}
s_{\mu_0^t}(x)
\prod_{a=1}^N 
\biggl[
s_{\mu_{a-1}/\eta_a}(q^{-\rho}) s_{\mu_{m_a}/\eta_a}(q^{-\rho})  (-Q_{m_a})^{|\mu_{m_a}|}  
s_{\mu_a^t/\sigma_a}(q^{-\rho})  s_{\mu_{m_a}^t/\sigma_a} (q^{-\rho})   (-Q_{a} )^{|\mu_a|}
\biggr]
s_{\mu_N}(y),
\label{pf_chain_2}
\ea
\ee
where $x=\{x_i\}$ and $y=\{y_i\}$ are the set of the eigenvalues of the holonomy matrices $V$ and $W$, respectively.
To compute this partition function, we apply the vertex operator formalism in the following.

The skew-Schur function is given as a matrix element of the vertex operator
\be
\ba
&s_{\mu/\nu}(x)=\langle\nu |V_+(x)|\mu\rangle = \langle\mu |V_-(x)|\nu\rangle,
~
&s_{\mu^t/\nu^t}(x)=\langle\nu |V'_+(x)|\mu\rangle = \langle\mu |V'_-(x)|\nu\rangle,
\ea
\ee
where $|\mu\rangle$ is the fermionic Fock state labelled by the partition $\mu = (\mu_1 \ge \mu_2 \ge \cdots \ge 0) \in \mathbb{Z}_{\ge 0}^\infty$, and $V_\pm (x), V'_\pm(x)$ are the vertex operators. Their concrete definitions are summarized in Appendix \ref{appA}. Then the partition function \eqref{pf_chain_2} is expressed as a chiral correlator of these vertex operators on a plane $\mathbb{C}$,
\be
\ba
\mathcal{Z}_{\text{chain}}^N
=
&\sum_{\text{all indices}}
\langle 0 | V'_+ (x) | \mu_0 \rangle
\\ &\hspace{1mm}
\times
\prod_{a=1}^N 
\Bigl[ 
\langle \mu_{a-1} |  V_- (q^{-\rho}) | \eta_a \rangle  \langle \eta_a | V_+(q^{-\rho}) | \mu_{m_a} \rangle   (-Q_{m_a})^{|\mu_{m_a}|}  
\langle \mu_{m_a} |  V'_- (q^{-\rho}) | \sigma_a \rangle  \langle \sigma_a | V'_+(q^{-\rho}) | \mu_{a} \rangle    (-Q_{a} )^{|\mu_a|}  
\Bigr]
 \\ &\hspace{1mm}
\times
\langle \mu_N | V_-(y) | 0 \rangle
\\
=&
\langle 0 |V'_+ (x)   \prod_{a=1}^N  \biggl[ V_- (q^{-\rho})   V_+(q^{-\rho})    (-Q_{m_a})^{L_0}  
 V'_- (q^{-\rho}) V'_+(q^{-\rho})  (-Q_{a} )^{L_0}  \biggr]  V_-(y)  | 0 \rangle
 \label{pf_chain_N}.
\ea
\ee
Note that the partition function \eqref{pf_chain_N} is reduced to the closed sting partition function on the $N$-chain geometry, when we remove the operators $V_+(x)$ and $V'_-(y)$, or equivalently set $x_i=y_i=0$ for $\forall i \in \mathbb{Z}_{>0}$,
\be
\ba
\mathcal{Z}_{\text{chain}}^{N,\text{c}}
=
\langle 0 | \prod_{a=1}^N  \biggl[ V_- (q^{-\rho})   V_+(q^{-\rho})    (-Q_{m_a})^{L_0}  
 V'_- (q^{-\rho}) V'_+(q^{-\rho})  (-Q_{a} )^{L_0}  \biggr]| 0 \rangle .
 \label{pf_chain_Nc}
\ea
\ee
Then we obtain the partition function with the formula~\eqref{VO_comm},
\be
\ba
\mathcal{Z}_{\text{chain}}^N
&=
 \prod_{a=1}^N \prod_{i,j=1}^\infty (1-Q_{m_a}q^{i+j-1})
 \prod_{a<b}^N \prod_{i,j=1}^\infty 
\frac{(1-Q_a \prod_{k=a+1}^{b-1} Q_{\tau_k} q^{i+j-1})
(1-Q_{m_b} \prod_{k=a}^{b-1} Q_{\tau_k} q^{i+j-1})}
{(1-Q_{a}Q_{m_b} \prod_{k=a+1}^{b-1} Q_{\tau_k} q^{i+j-1})
(1- \prod_{k=a}^{b-1} Q_{\tau_k} q^{i+j-1})}
 \nonumber \\
 &\quad\times
\prod_{a=1}^N \prod_{i,j=1}^\infty \frac
{(1-x_i \prod_{k=1}^{a-1}Q_{\tau_k}  q^{j-1/2})(1-y_i \prod_{k=a}^{N-1} Q_{\tau_k}  q^{j-1/2})}
{(1- x_i Q_{m_a}\prod_{k=1}^{a-1}Q_{\tau_k} q^{j-1/2})(1-y_i Q_{m_N} \prod_{k=a}^{N-1} Q_{\tau_k}  q^{j-1/2})}
\prod_{i,j=1}^\infty (1-x_i y_j Q_{m_N}\prod_{k=1}^{N-1} Q_{\tau_k}),
\ea
\ee
where $Q_{\tau_a}:=Q_{m_a}Q_a$, and we shift the variables $x_i\to-x_i$ and $y_i$, $Q_N y_i \to y_i$.
The closed string partition function is given by setting $x_i=y_i=0$ for $\forall i \in \mathbb{Z}_{>0}$,
\be
\ba
\mathcal{Z}_{\text{chain}}^{N,\text{c}}
=
 \prod_{a=1}^N \prod_{i,j=1}^\infty (1-Q_{m_a}q^{i+j-1})
 \prod_{a<b}^N \prod_{i,j=1}^\infty 
\frac{(1-Q_a \prod_{k=a+1}^{b-1} Q_{\tau_k} q^{i+j-1})
(1-Q_{m_b} \prod_{k=a}^{b-1} Q_{\tau_k} q^{i+j-1})}
{(1-Q_{a}Q_{m_b} \prod_{k=a+1}^{b-1} Q_{\tau_k} q^{i+j-1})
(1- \prod_{k=a}^{b-1} Q_{\tau_k} q^{i+j-1})}.
\label{pf_chain_Nc2}
\ea
\ee
Therefore the open string sector of the partition function $\mathcal{Z}_{\text{chain}}^{N,\text{o}}$ is obtained by subtracting the closed string contribution,
\be
\ba
\mathcal{Z}_{\text{chain}}^{N,\text{o}}
&:=
\frac{\mathcal{Z}^N_{\text{chain}}}{\mathcal{Z}_{\text{chain}}^{N,\text{c}}}
\\
&=
\prod_{a=1}^N \prod_{i,j=1}^\infty \frac
{(1-x_i  \prod_{k=1}^{a-1}Q_{\tau_k} q^{j-1/2})(1- y_i \prod_{k=a}^{N-1} Q_{\tau_k} q^{j-1/2})}
{(1- x_i Q_{m_a}\prod_{k=1}^{a-1}Q_{\tau_k} q^{j-1/2})(1- y_i  Q_{m_N} \prod_{k=a}^{N-1} Q_{\tau_k}q^{j-1/2})}
\prod_{i,j=1}^\infty (1-x_i y_j Q_{m_N}\prod_{k=1}^{N-1} Q_{\tau_k}).
\label{pf_chain_No}
\ea
\ee

Next we consider the $N$-chain geometry with the additional branes on the vertical lines as shown in Fig.~\ref{chain} (b).
The partition function associated with this geometry is similarly computed with the topological vertex formalism,
\be
\ba
{\mathcal{Z}'}_{\text{chain}}^N
&=
\sum_{\text{all indices}}
\prod_{a=1}^N \biggl[
(-Q_{m_a})^{|\mu_{m_a}|}(-Q_{a})^{|\mu_{a}|}
C_{\mu_{a-1}\mu_{m_a}\nu_a}(q)
C_{\mu_{a}^t \mu_{m_a}^t \nu'_a}(q)
{\rm Tr}_{\nu_a}U^{(a)} \hspace{1mm} {\rm Tr}_{{\nu'}_a} W^{(a)}
\biggr]
\\
&=
\sum_{\text{all indices}}
\prod_{a=1}^N \biggl[
q^{\frac{\kappa_{\nu_a}+\kappa_{{\nu'}^t_a}}{2}}s_{\nu_a}(q^{-\rho})s_{\nu'^t_a}(q^{-\rho}){\rm Tr}_{\nu_a}U^{(a)} \hspace{1mm} {\rm Tr}_{{\nu'}^t_a} W^{(a)}
\\
&\qquad\times
s_{\mu_{a-1}^t/\eta_a^t}(q^{-\rho-\nu_a})s_{\mu_{m_{a}}^t/\eta_a^t}(q^{-\rho-\nu^t_a})(-Q_{m_a})^{|\mu_{m_a}|}
s_{\mu_{a}/\sigma_a}(q^{-\rho-\nu'^t_a})s_{\mu_{m_{a}}/\sigma_a}(q^{-\rho-\nu'_a})(-Q_{a})^{|\mu_{a}|}
\biggr],
\label{pf_chain_3}
\ea
\ee
where we impose $\mu_0=\mu_N=\emptyset$. Again we can calculate the partition function by using the operator formalism,
\be
\ba
{\mathcal{Z}'}_{\text{chain}}^N
&=
\sum_{\text{all indices}}
\prod_{a=1}^N \biggl[
q^{\frac{\kappa_{\nu_a}+\kappa_{{\nu'}^t_a}}{2}}s_{\nu_a}(q^{-\rho})s_{\nu'^t_a}(q^{-\rho}){\rm Tr}_{\nu_a}U^{(a)} \hspace{1mm} {\rm Tr}_{{\nu'}^t_a} W^{(a)}
\biggr]
\\
&\qquad\times
\langle 0 |
\prod_{a=1}^N V_{-}(q^{-\rho-\nu_a}) V_{+}(q^{-\rho -\nu_a^t}) (-Q_{m_a})^{L_0} V'_{-}(q^{-\rho-\nu'_a}) V'_{+}(q^{-\rho -{\nu'}_a^t}) (-Q_{a})^{L_0} 
|0 \rangle,
\ea
\ee
and the result is
\be
\ba
{\mathcal{Z}'}_{\text{chain}}^N
=
&\sum_{\text{all indices}}
\prod_{a=1}^N \biggl[q^{\frac{\kappa_{\nu_a}+\kappa_{{\nu'}^t_a}}{2}}s_{\nu_a}(q^{-\rho})s_{\nu'^t_a}(q^{-\rho}){\rm Tr}_{\nu_a}U^{(a)} \hspace{1mm} {\rm Tr}_{{\nu'}^t_a} W^{(a)}\biggr]
\\
&\times
\prod_{a=1}^N \prod_{i,j=1}^\infty(1-Q_{m_a} q^{i+j-\nu_{a,j}^t -\nu'_{a,i}-1})
\\
&\times
\prod_{a<b}^{N}
\frac{(1-Q_a \prod_{k=a+1}^{b-1}Q_{\tau_k}q^{i+j-{\nu'}_{a,j}^t-\nu_{b,i}-1})(1-Q_{m_b} \prod_{k=a+1}^{b-1}Q_{\tau_k}q^{i+j-{\nu^t}_{a,j}-{\nu'}_{b,i}-1})}
{(1-\prod_{k=a}^{b-1}Q_{\tau_k}q^{i+j-{\nu}_{b,i}-\nu_{a,j}^t-1})(1-Q_{m_b} Q_a\prod_{k=a+1}^{b-1}Q_{\tau_k}q^{i+j-{\nu'}^t_{a,i}-{\nu'}_{b,j}-1})},
\ea
\ee
so that the brane contribution is given by
\be
\ba
{\mathcal{Z}'}^{N,\text{o}}_{\text{chain}}
&:=
\frac{{\mathcal{Z}'}^N_{\text{chain}}}{\mathcal{Z}_{\text{chain}}^{N,\text{c}}}
\\
&=
\sum_{\text{all indices}}
\prod_{a=1}^N \biggl[q^{\frac{||\nu_a^t||^2}{2}}s_{\nu_a}(x^{(a)}) \hspace{1mm}q^{\frac{||\nu_a||^2}{2}} s_{\nu'^t_a}(y^{(a)})\biggr]
\\
&\quad\times
\prod_{a=1}^N 
\prod_{(i,j)\in \nu_a}\frac{(1-Q_{m_a} q^{-i-j+\nu_{a,i}+{\nu'}^t_{a,j}+1})}{(1-q^{-i-j+\nu_{a,i} +\nu_{a,j}^t +1})}
\prod_{(i,j)\in \nu'_a}\frac{(1-Q_{m_a} q^{i+j-\nu_{a,j}^t -\nu'_{a,i}-1})}{(1-q^{i+j-{\nu'}_{a,i} -{\nu'}_{a,j}^t -1})}
\\
&\quad\times
\prod_{a<b}^N
\prod_{(i,j)\in \nu_a}
\frac{(1-Q_{m_b} \prod_{k=a}^{b-1}Q_{\tau_k}q^{-i-j+{\nu}_{a,i}+{\nu'}^t_{b,j}+1})}{(1-\prod_{k=a}^{b-1}Q_{\tau_k}q^{-i-j+{\nu}_{a,i}+{\nu}_{b,j}^t+1})}
\prod_{(i,j)\in \nu'_a}
\frac{(1-Q_{a} \prod_{k=a+1}^{b-1}Q_{\tau_k}q^{-i-j+{\nu'}_{a,i}+{\nu}_{b,j}^t+1})}{(1-Q_{m_b}Q_a \prod_{k=a+1}^{b-1}Q_{\tau_k}q^{-i-j+{\nu'}_{a,i}+{\nu'}_{b,j}^t+1})}
\\
&\hspace{9.5mm}\times
\prod_{(i,j)\in \nu_b}
\frac{(1-Q_{a} \prod_{k=a+1}^{b-1}Q_{\tau_k}q^{i+j-{\nu}_{b,i}-{\nu'}^t_{a,j}-1})}{(1-\prod_{k=a}^{b-1}Q_{\tau_k}q^{i+j-{\nu}_{b,i}-{\nu}_{b,j}^t+1})}
\prod_{(i,j)\in \nu'_b}
\frac{(1-Q_{m_b} \prod_{k=a}^{b-1}Q_{\tau_k}q^{i+j-{\nu'}_{b,i}-{\nu}_{a,j}^t-1})}{(1-Q_{m_b}Q_a \prod_{k=a+1}^{b-1}Q_{\tau_k}q^{i+j-{\nu'}_{b,i}-{\nu}_{a,j}^t-1})},
\ea
\ee
where $x^{(a)}$ and $y^{(a)}$ are the sets of the eigenvalues of the holonomy matrices $U^{(a)}$ and $W^{(a)}$, respectively.

Note that, if we restrict the geometry to the single brane insertion, namely $x^{(a)}_i=\delta_{a1}\delta_{i1} x$ and $y^{(a)}_i=0$, the partition function can be expressed as the basic $q$-hypergeometric function\footnote{In order to be simple expression, we shift $Q_a \to q Q_a $ for $a=1,...,N-1$ and $Q_{m_a} \to q^{-1} Q_{m_a}$ for $a=2,...,N$.},
\be
\ba
{\mathcal{Z}'}^{N,\text{o}}_{\text{chain}}
&=
\sum_{n=0}^\infty
(q^{1/2}x)^n
\prod_{a=1}^N
\frac{(Q_{m_a}\prod_{k=1}^{a-1} Q_{\tau_k}; q)_n }{(q \prod_{k=1}^{a-1} Q_{\tau_k}; q)_n}
\\
&=: \vspace{0.1mm}_N \phi_{N-1} (Q_{m_1},Q_{m_2}Q_{\tau_1},Q_{m_3}Q_{\tau_1}Q_{\tau_2},\cdots Q_{m_N}\prod_{k=1}^{N-1}Q_{\tau_k};Q_{\tau_1},Q_{\tau_1}Q_{\tau_2},\cdots,\prod_{k=1}^{N-1}Q_{\tau_k};q;q^{1/2}x),
\label{pf_chain_No2}
\ea
\ee
where the basic $q$-hypergeometric function $ \vspace{0.1mm}_r \phi_{s-1}(a_1,a_2,...,a_r; b_1,b_2,...,b_s ; q; z)$ and $q$-shifted factorial ($q$-Pochhammer symbol) $(x;q)_n~(n\geq 1) $ are defined as
\be
\ba
& \vspace{0.1mm}_r \phi_{s-1}(a_1,a_2,...,a_r; b_1,b_2,...,b_s ; q; z)
\\
&\qquad =
 \sum_{n=0}^\infty
  \frac{z^n}{(q;q)_n}\bigl( (-1)^n q^{1/2 n(n-1)} \bigr)^{1+s-r}
 \frac{(a_1;q)_n (a_2;q)_n \cdots (a_r;q)_n}{(b_1;q)_n (b_2;q)_n \cdots (b_s;q)_n},
 \\
&\quad
(x;q)_n = \prod_{k=0}^{n-1}(1-x q^k).
\ea
\ee
The $q$-difference equation for this $q$-hypergeometric partition function will play a key role in quantization of the mirror curve, as discussed in Section~\ref{mirror_curve}.

\subsection{Compactification}
\begin{figure}[htb]
\begin{minipage}{0.5\hsize}
\includegraphics[width=8cm]{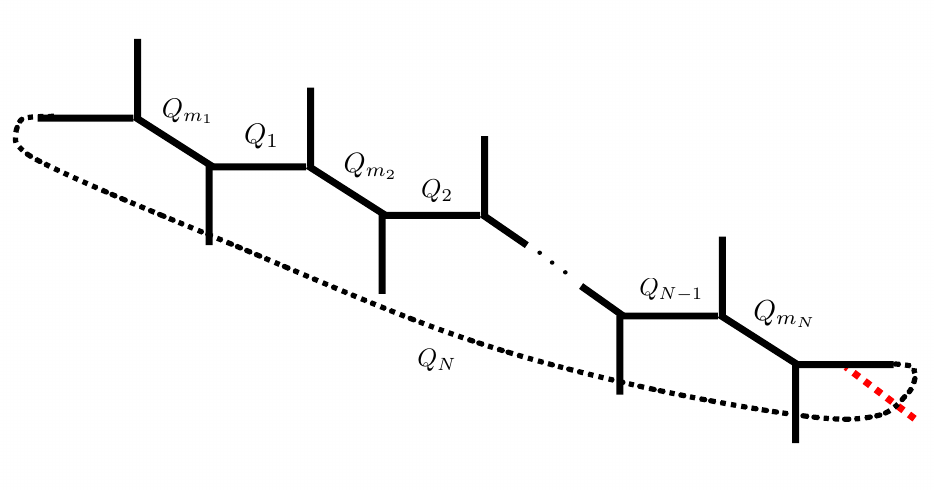}
\end{minipage}
\put(-225,65){(a)}
\hspace{8mm}
\begin{minipage}{0.45\hsize}
\includegraphics[width=8cm]{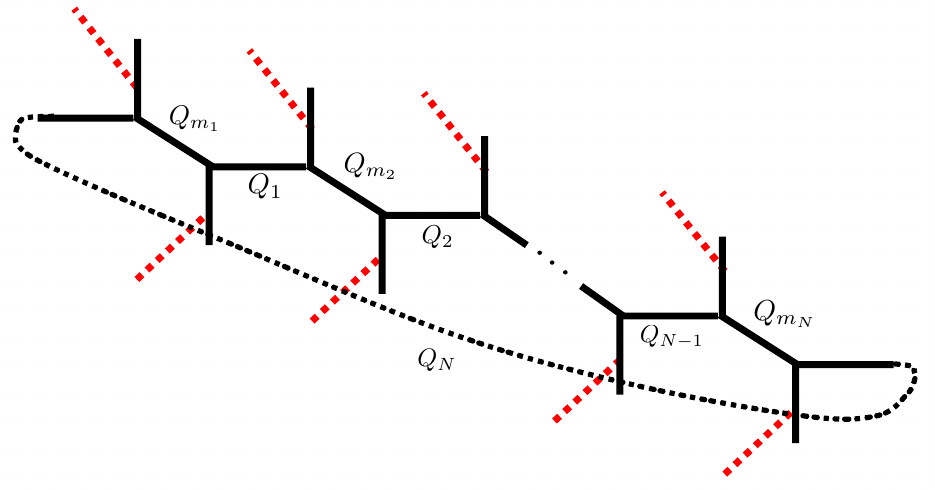}
\end{minipage}
\put(-205,65){(b)}
\caption{The periodic $N$-chain geometry with the branes. Compared with Fig.~\ref{chain}, there is compactified direction along the horizontal line.}
\label{period_chain}
\end{figure}

Next we calculate the partition function of the topological strings on periodic $N$-chain geometries defined in Fig.~\ref{period_chain}. 

\begin{figure}[htb]
\centering
\includegraphics[width=6cm]{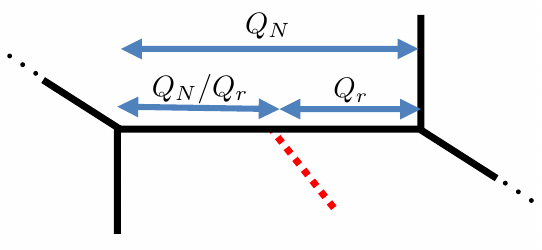}
\caption{A part of the web diagram in Fig.~\ref{period_chain} (a).}
\label{inner_brane}
\end{figure}

To begin with, we calculate the partition function in Fig.~\ref{period_chain} (a) which is given by
\be
\ba
\mathcal{Z}^N_{\text{period}}
&=
\sum_{\text{all indices}}
(-Q_{m_1})^{|\mu_{m_1}|} (-Q_{1} )^{|\mu_1|}
 (-1)^{p|\nu_1|}(Q_N/Q_r)^{|\nu_1|}q^{\frac{p}{2}\kappa_{\mu_0^t \otimes\nu_1}}
C_{\mu_0^t \otimes \hspace{0.3mm}\nu_1 \mu_{m_1} \emptyset}  C_{\mu_1 \mu^t_{m_{1}}\emptyset} {\rm Tr}_{\nu_1}U
\\&\quad\times
\prod_{a=2}^{N-1} \Bigl[ 
(-Q_{m_a})^{|\mu_{m_a}|} (-Q_{a} )^{|\mu_a|}
C_{\mu_{a-1}^t \mu_{m_a} \emptyset}  C_{\mu_a \mu^t_{m_{a}} \emptyset}  \Bigr]
\\&\quad\times
 (-Q_{m_N})^{|\mu_{m_N}|}(-Q_{N})^{|\mu_N|} C_{\mu_{N-1}^t \mu_{m_N} \emptyset} 
(-1)^{p|\nu_2|}Q_r^{|\nu_2|}  q^{\frac{p}{2}\kappa_{\mu_N\otimes\nu_2}}
C_{\mu_N\otimes\nu_2 \hspace{0.3mm} \mu_{m_N}\emptyset} {\rm Tr}_{\nu_2}U^{-1},
\ea
\ee
where  $p$ is the framing of the branes that we will set $p=0$, and the parameter $Q_r$ corresponds to the location of the branes, as in Fig.~\ref{inner_brane}.
By using the following formula for the topological vertex \cite{Awata:2010bz},
\be
\ba
C_{\alpha\otimes\beta \mu\nu}
=
\sum_{\gamma}N^{\gamma}_{\alpha\beta}C_{\gamma\mu\nu},
\label{vertex_formula}
\ea
\ee
where $N^{\alpha}_{\mu\nu}$ is the Littlewood--Richardson coefficient, we find
\be
\ba
\mathcal{Z}^N_{\text{period}}
&=
\sum_{\text{all indices}}
(-Q_{m_1})^{|\mu_{m_1}|}(-Q_1)^{|\mu_1|} s_{\gamma/\mu_0^t}(x_1) s_{\gamma^t/\eta_1}(q^{-\rho})s_{\mu_{m_1}/\eta_1}(q^{-\rho})s_{\mu_1^t/\sigma_1}(q^{-\rho})s_{\mu_{m_1}^t/\sigma_1}(q^{-\rho})
\\
&\quad\times
\prod_{a=2}^{N-1}\biggl[(-Q_{m_a})^{|\mu_{m_a}|} (-Q_{a})^{|\mu_a|} 
s_{\mu_{a-1}/\eta_a}(q^{-\rho}) s_{\mu_{m_a}/\eta_a}(q^{-\rho}) s_{\mu_{a}^t/\sigma_a}(q^{-\rho}) s_{\mu_{m_a}^t/\sigma_a}(q^{-\rho})
\biggr]
\\
&\quad\times
s_{\mu_{N-1}/\eta_N}(q^{-\rho}) s_{\mu_{m_N}/\eta_N}(q^{-\rho}) s_{\delta^t/\sigma_N}(q^{-\rho})s_{\mu_{m_N}^t/\sigma_N}(q^{-\rho})
 s_{\delta/\mu_N}(x_2),
\ea
\ee
where $x_{1,i}:=x_{i}\frac{Q_{N,1}}{Q_r},~x_{2,i}:=\frac{1}{x_i}Q_r$. In the operator formalism, the partition function can be expressed as a torus correlator,
\be
\ba
\mathcal{Z}_{\text{period}}^N
&=
{\rm Tr}\biggl[
V'_+(x_1)
\prod_{a=1}^N  \biggl[ V_- (q^{-\rho})   V_+(q^{-\rho})    (-Q_{m_a})^{L_0}  
 V'_- (q^{-\rho}) V'_+(q^{-\rho})  (-Q_{a} )^{L_0}  \biggr] 
V_-(-Q_N^{-1}x_2)
\biggr].
\ea
\ee
At this point, we find that the compactification implies to replace the expectation value $\langle 0 | \cdots | 0 \rangle$ in the chain geometry with the trace ${\rm Tr}[\cdots]$ with $x=x_{1},~y=x_2$.
The closed string partition function is thus given by
\be
\ba
\mathcal{Z}_{\text{period}}^{N,c}
&=
{\rm Tr}
\biggl[
\prod_{a=1}^N  \biggl[ V_- (q^{-\rho})   V_+(q^{-\rho})    (-Q_{m_a})^{L_0}  
 V'_- (q^{-\rho}) V'_+(q^{-\rho})  (-Q_{a} )^{L_0}  \biggr] 
\biggr]
\ea
\ee
which is the $2N$-point correlation function of free fermions on a torus.

By using the trace formula%
\footnote{See, for example, \cite{Kimura:2016dys} for the derivation.},
\be
\ba
{\rm Tr}\bigl[a^{x\partial_x}{\rm e}^{bx}{\rm e}^{c\partial_x}\bigr]=\frac{1}{1-a}{\rm exp}\biggl[\frac{abc}{1-a}\biggr],
\label{trace}
\ea
\ee
we find the following result,
\be
\ba
\mathcal{Z}_{\text{period}}^N
&=
\prod_{n=1}^{\infty}
\frac{1}{(1-Q_\tau^n)}
\prod_{a,b=1}^{N}
\prod_{i,j=1}^\infty
\frac{
(1- Q_{ab} q^{i+j-1} Q_\tau ^{n-1})(1- Q_{ab}^{-1}  q^{i+-1}Q_\tau^{n})
}{
(1- \tilde{Q}'_{ab} q^{i+j-1}  Q_\tau^{n-1})(1- \tilde{Q}_{ab} q^{i+j-1} Q_\tau^{n-1})
}
\\
&\quad\times
\prod_{n=1}^{\infty} \prod_{i,j=1}^\infty
(1-x_{i}x_{j}^{-1} Q_{\tau}^{n})
\\
&\quad\times
\prod_{n=1}^{\infty}
\prod_{a=1}^N \prod_{i,j=1}^\infty
\frac{
(1-x_{i}^{-1} Q_r^{-1}Q_{m_a}^{-1} \prod_{k=a}^N Q_{\tau_k}  q^{j-1/2} Q_{\tau}^{n-1})(1- x_{i} Q_r  \prod_{k=a}^N Q_{\tau_k}^{-1}q^{j-1/2} Q_{\tau}^{n})
}{
(1-x_{i}^{-1}Q_r^{-1} \prod_{k=a}^N Q_{\tau_k} q^{j-1/2} Q_{\tau}^{n-1})(1-x_{i}Q_r Q_{m_a}  \prod_{k=a}^N Q_{\tau_k}^{-1} q^{j-1/2} Q_{\tau}^{n})
},
\label{pf_period1}
\ea
\ee
where we define the following variables,
\be
\ba%
&Q_\tau = \prod_{k=1} ^N Q_{\tau_k},
\\
&Q_{ab}
=
\begin{cases}
     Q_{m_a}\prod_{k = b}^{N} Q_{\tau_k} \quad
     (\text{mod } Q_\tau) \quad
     \text{for } a = 1, \\ 
     Q_{m_a} \prod_{k = 1}^{a - 1} Q_{\tau_k} \prod_{l = b}^{N}Q_{\tau_l} \quad
     (\text{mod } Q_\tau) \quad
    \text{for } a \neq 1, 
\end{cases}
\\
&\tilde{Q}_{ab}
=
\begin{cases}
 \prod_{k = b}^{a - 1} Q_{\tau_k} \quad
   \text{for } a >b, \\ 
  Q_\tau  \quad
   \text{for } a = b, \\ 
   Q_\tau / \prod_{k = a}^{b - 1} Q_{\tau_k} \quad
   \text{for } a < b, 
\end{cases}
\\
&\tilde{Q}'_{ab} = \frac{Q_{m_a}}{Q_{m_b}}\tilde{Q}_{ab},
\label{def}
\ea
\ee
and we shift the variables, $x_i\to-x_i$ and $y_i\to -y_i$.
The first line in \eqref{pf_period1} is just the closed string partition function $\mathcal{Z}^{\text{c}}_{N,\text{period}}$, so that the remaining contribution is the partition function of the open topological string,
\be
\ba
\mathcal{Z}_{\text{period}}^{N,\text{o}}
&=
\frac{\mathcal{Z}^N_{\text{period}}}{\mathcal{Z}_{\text{period}}^{N,\text{c}}}
\\
&=
\prod_{n=1}^{\infty} \prod_{i,j=1}^\infty
(1-x_{i}x_{j}^{-1} Q_{\tau}^{n})
\prod_{a=1}^N 
\frac{
(1-Q_{m_a}^{-1} x_{i}^{-1} \prod_{k=a}^N Q_{\tau_k}  q^{j-1/2} Q_{\tau}^{n-1})(1- x_{i} \prod_{k=a}^N Q_{\tau_k}^{-1}  q^{j-1/2} Q_{\tau}^{n})
}{
(1-x_{i}^{-1} \prod_{k=a}^N Q_{\tau_k} q^{j-1/2} Q_{\tau}^{n-1})(1-Q_{m_a} x_{i}  \prod_{k=a}^N Q_{\tau_k}^{-1} q^{j-1/2} Q_{\tau}^{n})
}.
\label{pf_period_No}
\ea
\ee
To obtain the simple expression, we shift the variables, $x_i \to x_i/Q_r$.

Finally we calculate the partition function of the topological string on the toric Calabi--Yau threefold in Fig.~\ref{period_chain} (b) which has been calculated in \cite{Haghighat:2013gba, Haghighat:2013tka, Yoshida:2014qwa}.
\be
\ba
\mathcal{Z'}_{\text{period}}^N
&=
\sum_{\text{all indices}}
\prod_{a=1}^N \biggl[
(-Q_{m_a})^{|\mu_{m_a}|}(-Q_{a})^{|\mu_{a}|}
C_{\mu_{a-1}\mu_{m_a}\nu_a}(q)
C_{\mu_{a}^t \mu_{m_a}^t \nu'_a}(q)
{\rm Tr}_{\nu_a}U^{(a)} \hspace{1mm} {\rm Tr}_{{\nu'}_a} W^{(a)}
\biggr]
\\
&=
\sum_{\text{all indices}}
\prod_{a=1}^N \biggl[
q^{\frac{\kappa_{\nu_a}+\kappa_{{\nu'}^t_a}}{2}}s_{\nu_a}(q^{-\rho})s_{\nu'^t_a}(q^{-\rho}){\rm Tr}_{\nu_a}U^{(a)} \hspace{1mm} {\rm Tr}_{{\nu'}^t_a} W^{(a)}
\\
&\qquad\times
s_{\mu_{a-1}^t/\eta_a^t}(q^{-\rho-\nu_a})s_{\mu_{m_{a}}^t/\eta_a^t}(q^{-\rho-\nu^t_a})(-Q_{m_a})^{|\mu_{m_a}|}
s_{\mu_{a}/\sigma_a}(q^{-\rho-\nu'^t_a})s_{\mu_{m_{a}}/\sigma_a}(q^{-\rho-\nu'_a})(-Q_{a})^{|\mu_{a}|}
\biggr],
\label{pf_period}
\ea
\ee
where we impose the periodic boundary condition, $\mu_N=\mu_0$. 
%
After the similar computation, we find 
\be 
\ba
\mathcal{Z'}_{\text{period}}^N
&=
\sum_{\text{all indices}}
\prod_{a=1}^N \biggl[
q^{\frac{\kappa_{\nu_a}+\kappa_{{\nu'}^t_a}}{2}}s_{\nu_a}(q^{-\rho})s_{\nu'^t_a}(q^{-\rho}){\rm Tr}_{\nu_a}U^{(a)} \hspace{1mm} {\rm Tr}_{{\nu'}^t_a} W^{(a)}
\biggr]
\\&\quad\times
\prod_{n=1}^{\infty}
\frac{1}{(1-Q_\tau^n)}
\prod_{a,b=1}^{N}
\prod_{i,j=1}^\infty
\frac{
(1- Q_{ab} q^{i+j-{\nu'}_{b,i}+{\nu}^t_{a,j}-1} Q_\tau ^{n-1})(1- Q_{ab}^{-1}  q^{i+j-{\nu}_{a,i}-{\nu'}^t_{b,j}-1}Q_\tau^{n})
}{
(1- \tilde{Q}_{ab} q^{i+j-{\nu}_{b,i}-{\nu}^t_{a,j}-1}  Q_\tau^{n-1})(1- \tilde{Q}'_{ab} q^{i+j-{\nu'}_{a,i}-{\nu'}^t_{b,j}-1} Q_\tau^{n-1})
}.
\ea
\ee
Dividing the full partition function $\mathcal{Z'}_{\text{period}}^N$ by the closed string contribution $\mathcal{Z}_{\text{period}}^{\text{c},N}$, we obtain the partition function of the open topological string,
\be
\ba
\mathcal{Z'}_{\text{period}}^{N,\text{o}}
&=
\frac{\mathcal{Z'}^N_{\text{period}}}{\mathcal{Z}_{\text{period}}^{N,\text{c}}}
\\
&=
\sum_{\text{all indices}}
\prod_{a=1}^N \biggl[
q^{\frac{\kappa_{\nu_a}+\kappa_{{\nu'}^t_a}}{2}}s_{\nu_a}(q^{-\rho})s_{\nu'^t_a}(q^{-\rho})s_{\nu_a}(x^{(a)}) s_{{\nu'}^t_a} (y^{(a)})
\biggr]
\\&\quad\times
\prod_{n=1}^{\infty}
\prod_{a,b=1}^{N}
\prod_{(i,j)\in\nu_a}
\frac{(1- Q_{ab} q^{-i-j+{\nu}_{a,i}+{\nu'}_{b,j}^t+1} Q_\tau ^{n-1})(1- Q_{ab}^{-1}  q^{i+j-{\nu}_{a,i}-{\nu'}^t_{b,j}-1}Q_\tau^{n})
}{(1- \tilde{Q}_{ab} q^{-i-j+{\nu}_{a,i}+{\nu}_{b,j}^t+1}  Q_\tau^{n-1})(1- \tilde{Q}_{ba} q^{i+j-{\nu}_{a,i}-{\nu}^t_{b,j}-1}  Q_\tau^{n-1})}
\\&\qquad\qquad\quad\times
\prod_{(i,j)\in\nu'_b}
\frac{
(1- Q_{ab} q^{i+j-{\nu'}_{b,i}+{\nu}^t_{a,j}-1} Q_\tau ^{n-1})(1- Q_{ab}^{-1}q^{-i-j+{\nu'}_{a,i}+{\nu}_{b,j}^t +1}Q_\tau^{n})}
{(1- \tilde{Q}'_{ab} q^{-i-j+{\nu'}_{a,i}+{\nu'}_{b,i}^t+1} Q_\tau^{n-1})(1- \tilde{Q}'_{ba} q^{i+j-{\nu'}_{b,i}-{\nu'}^t_{a,j}-1} Q_\tau^{n-1})}.
\ea
\ee

Again, for the single brane geometry,  $x^{(a)}_i=\delta_{a1}\delta_{i1} x$ and $y^{(a)}_i=0$, the partition function of the open topological strings reduces to the elliptized hypergeometric function, called ``elliptic hypergeometric function,''\footnote{Again, we shift $Q_a \to q Q_a $ for $a=1,...,N-1$ and $Q_{m_a} \to q^{-1} Q_{m_a}$ for $a=2,..,N$.}
\be
\ba
\mathcal{Z'}_{\text{period}}^{N,\text{o}}
&=
\sum_{n=0}^\infty
(q^{1/2}x)^n
\prod_{a=1}^N
\frac{\theta(Q_{m_a}^{-1}\prod_{k=1}^{a-1} Q^{-1}_{\tau_k};q^{-1}; Q_\tau)_n }{\theta(q^{-1}\prod_{k=1}^{a-1} Q^{-1}_{\tau_k};q^{-1}; Q_\tau)_n}
\\
&= \vspace{0.1mm}_N E_{N-1} (Q_{m_1}^{-1},Q_{m_2}^{-1}Q^{-1}_{\tau_1},\cdots Q_{m_N}^{-1}\prod_{k=1}^{N-1}Q^{-1}_{\tau_k};Q^{-1}_{\tau_1},Q^{-1}_{\tau_1}Q^{-1}_{\tau_2},\cdots,\prod_{k=1}^{N-1}Q^{-1}_{\tau_k};q^{-1};q^{1/2}x),
\label{pf_period_No2}
\ea
\ee
where we define the theta function\footnote{The definition is somewhat unusual, however, we use this definition in order to match to that in some papers about M-strings, for example, \cite{Haghighat:2013gba}.}
\begin{align}
 \theta(z;p) = (z^{-1};p)_\infty (pz;p)_\infty ,
\end{align}
and the elliptic hypergeometric function and elliptization of the $q$-Pochhammer symbol are defined as
\be
\ba
& _r E_{r-1} (a_1,a_2,...,a_r;b_1,b_2,...,b_{r-1};z)
\\
&\quad
=
\sum_{n=0}^\infty
z^n
\frac{\theta(a_1 ;q ;p)_n \theta(a_2 ;q ;p)_n \cdots \theta(a_r ;q ;p)_n}
{\theta(q;q ;p)_n \theta(b_1 ;q ;p)_n \theta(b_2 ;q ;p)_n \cdots \theta(b_{r-1} ;q ;p)_n},
 \\
&\quad
\theta(a;q;p)_n = \prod_{i=0}^{n-1}\theta(a q^i;p).
\ea
\ee
We will discuss the $q$-difference equation for this hypergeometric function in Section~\ref{mirror_curve}.

\section{Mirror curve of chain geometry}\label{mirror_curve}
Now we shall derive the quantum operator $\mathsf{H}(\mathsf{x},\mathsf{y})$ from the partition function of the topological string with the single brane by the method in \cite{Taki:2010bj, Takasaki:2015raa}. The coupling constant $g_s$ plays a role of the Planck constant, so that we can obtain the classical mirror curve by taking the limit $g_s \to 0$.

The quantum operators annihilate the partition function of the open topological string, so that the task is how to find such operators. Here we present the procedure of finding $\mathsf{H}(\mathsf{x},\mathsf{y})$ and $H(x,y)$:
\begin{itemize}
\item[1.]
Regarding the partition function as a function of the modulus of the single brane $\mathcal{Z}(x)$, find the relation between $\mathcal{Z}(qx)$ and $\mathcal{Z}(x)$.
\item[2.]
Promoting the classical value $q$ to the operator $\mathsf{p}=\re^{\ri g_s \partial_u}$ where $u$ is related to the modulus $x=\re^u$, so that $\mathsf{p}x=qx$.
\item[3.]
Express the relation between $\mathcal{Z}(qx)$ and $\mathcal{Z}(x)$ as a difference equation, $\mathsf{H}(\mathsf{x}, \mathsf{p})\mathcal{Z}(x)=0$ which is just the quantum mirror curve. Then, we find the operator $\mathsf{H}(\mathsf{x}, \mathsf{p})$ and classical polynomial $H(x,p)$, and the equation $H(x,p)=0$ is the classical mirror curve $\Sigma$. The variable $q$ in the quantum operator corresponds to the ambiguity of the ordering of the operators, $\mathsf{x}$ and $\mathsf{p}$.
\end{itemize}

Let us demonstrate the above procedure in the simplest case, $N=1$, which corresponds to the resolved conifold and its compactified geometry. This is the special case since the geometry has the extra symmetry under exchanging the branes on the vertical and horiontal axis. After that, we will discuss the general case, $N > 1$.

\subsection{The simplest case $N=1$}\label{sec:N=1}
In the case of $N=1$, the partition functions of the topological strings are given from the results \eqref{pf_chain_No}, \eqref{pf_chain_No2}, \eqref{pf_period_No}, \eqref{pf_period_No2}, in the previous Section,
\begin{subequations}
\begin{align}
\mathcal{Z}_{\text{chain}}^{1,\text{o}}
&=
\frac{(q^{1/2} x ;q)_\infty}{(q^{1/2}xQ_{m} ;q)_\infty},
\\
{\mathcal{Z}'}^{1,\text{o}}_{\text{chain}} 
&=
\sum_{n=0}^\infty  (q^{1/2} x)^n  \frac{(Q_m;q)_n}{(q;q )_n},
\\
{\mathcal{Z}}^{1,\text{o}}_{\text{period}}
&=
\frac{\Gamma_e (q^{1/2} xQ_m;q;Q_\tau)}{\Gamma_e (q^{1/2} x;q;Q_\tau)},
\\
{\mathcal{Z}'}^{1,\text{o}}_{\text{period}}
&=
\sum_{n=0}^\infty
(q^{1/2}x)^n   \frac{\theta(Q_m^{-1};q^{-1};Q_\tau)_n}{\theta(q^{-1};q^{-1};Q_\tau)_n}
,
\end{align}
\end{subequations}
where we set $x_1 = x,~x_{i > 1}=0,~y_i=0$ in \eqref{pf_chain_No} and \eqref{pf_period_No}, and elliptic gamma function, $\Gamma_e(x,;p_1;p_2)$ is defined as
\be
\ba
\Gamma_e(x;p_1;p_2)
:=\prod_{m,n\geq0}\frac{1-x^{-1}p_1^{n+1}p_2^{m+1}}{1-x p_1^{n}p_2^{m}}.
\ea
\ee
To obtain ${\mathcal{Z}}^{1,\text{o}}_{\text{period}}$, we use following analytic continuation,
\be
\ba
\prod_{j=1}^\infty (1-Aq^{j-1}) = \prod_{j=1}^\infty (1-A q^{-j})^{-1}.
\ea
\ee
The former infinite product is for the region $|q|<1$, while the latter is for $|q|>1$.

Interestingly, from the $q$-binomial theorem,
\be
\ba
\sum_{n=0}^\infty \frac{(a;q)_n}{(q;q)_n}z^n = \frac{(a z;q)_\infty}{(z;q)_\infty},~(|q|<1,~|z|<1),
\ea
\ee
we have ${\mathcal{Z}}^{1,\text{o}}_{\text{chain}}(Q_m^{-1} x)\Big|_{Q_m \to Q_m^{-1}}={\mathcal{Z}'}^{1,\text{o}}_{\text{chain}}$, which is known as the flop transition \cite{Konishi:2006ev, Taki:2008hb}. The partition functions, ${\mathcal{Z}}^{1,\text{o}}_{\text{period}}$ and ${\mathcal{Z}'}^{1,\text{o}}_{\text{period}}$, are the elliptizations of the partition functions, ${\mathcal{Z}}^{1,\text{o}}_{\text{chain}}$ and ${\mathcal{Z}'}^{1,\text{o}}_{\text{chain}}$, respectively. It would be interesting to find the relation between ${\mathcal{Z}}^{1,\text{o}}_{\text{period}}$ and ${\mathcal{Z}'}^{1,\text{o}}_{\text{period}}$ as an extension of the $q$-binomial theorem.

In the chain geometry, by shifting the variable $x$ to $qx$, we find the following relation,
\be
\ba
\mathcal{Z}_{\text{chain}}^{1,\text{o}}(qx) = 
\frac{1-q^{1/2} Q_m x}{1-q^{1/2} x}
\mathcal{Z}_{\text{chain}}^{1,\text{o}}(x).
\ea
\ee
From this expression, the partition function satisfies the quantum mirror curve,
\be
\ba
\bigl[(1-q^{1/2} \mathsf{x} )\mathsf{p}-(1-q^{1/2} Q_m \mathsf{x} )\bigr]\mathcal{Z}_{\text{chain}}^{1,\text{o}}(x) =0.
\ea
\ee
Hence we find the operator $\mathsf{H}^1_{\text{chain}}$,
\be
\ba
\mathsf{H}^1_{\text{chain}}(\mathsf{x},\mathsf{p})=q^{1/2}Q_m \mathsf{x}+\mathsf{p}+q^{1/2}\mathsf{x}\mathsf{p}+1,
\ea
\ee
where we flip the sign of variables, $\mathsf{x}\to -\mathsf{x}$, $\mathsf{p}\to -\mathsf{p}$, and $\mathsf{H}_{\text{chain}} \to -\mathsf{H}_{\text{chain}}$. The classical mirror curve is
\be
\ba
H_{\text{chain}}^1(x,p) =Q_m x +p + xp +1=0,
\ea
\ee
which is well-known result for the resolved conifold.

We next derive the operator from ${\mathcal{Z}'}_{\text{chain}}^{1,\text{o}}(x)$. By defining the coefficient of $x^n$ as $\mathcal{Z}^{(n)}$,
\be
\ba
{\mathcal{Z}'}_{\text{chain}}^{1,\text{o}}(x)
 = \sum_{n=0}^\infty \mathcal{Z}^{(n)} \, x^n
\, , \qquad
\mathcal{Z}^{(n)} =  q^{n/2}  \frac{(Q_m; q)_n}{(q;q )_n},
\ea
\ee
we find the following difference equation,
\be
\ba
x^{n+1}\mathcal{Z}^{(n+1)} =q^{1/2}x \frac{(1-Q_m q^{n})}{(1-q^{n+1} )}   \times x^{n}\mathcal{Z}^{(n)},
\ea
\ee
so that the quantum mirror curve is
\be
\ba
\bigl[
(1-\mathsf{p})-q^{1/2}\mathsf{x}(1-Q_m\mathsf{p})
\bigr]
{\mathcal{Z}'}_{\text{chain}}^{1,\text{o}}(x)=0.
\ea
\ee
The operator ${\mathsf{H}'}^1_{\text{chain}}$ is then
\be
\ba
{\mathsf{H}'}^1_{\text{chain}}(\mathsf{x},\mathsf{p})=q^{1/2}\mathsf{x}+\mathsf{p}+q^{1/2}Q_m \mathsf{x}\mathsf{p}+1,
\ea
\ee
where we flip the sign of variables, $\mathsf{x}\to -\mathsf{x}$ and $\mathsf{p}\to -\mathsf{p}$. Therefore, we find
\be
\ba
\mathsf{H}^1_{\text{chain}}(Q_m^{-1}\mathsf{x},\mathsf{p})\Big|_{Q_m \to Q_m^{-1}}={\mathsf{H}'}^1_{\text{chain}}(\mathsf{x},\mathsf{p}).
\ea
\ee
 We should stress that this relation is satisfied only $N=1$ due to the geometrical symmetry. In Sec.~\ref{sec:N>1}, we will discuss the general case, and find the similar relation.

The similar calculation provides the quantum operators of the periodic $1$-chain geometries,
\begin{subequations}
\begin{align}
&\mathsf{H}^1_{\text{period}}(\mathsf{x},\mathsf{p})=\theta(q^{-1/2}Q_m^{-1}\mathsf{x}^{-1};Q_\tau)-\theta(q^{-1/2} \mathsf{x}^{-1};Q_\tau)\mathsf{p},
\\
&{\mathsf{H}'}^1_{\text{period}}(\mathsf{x},\mathsf{p})=\theta(\mathsf{p}^{-1};Q_\tau)-q^{1/2}\mathsf{x}\theta(Q_m^{-1}\mathsf{p}^{-1};Q_\tau),
\end{align}
\end{subequations}
and classical polynomials
\begin{subequations} 
 \begin{align}
  &H_{\text{period}}^1(x,p)=\theta(Q_m^{-1}x^{-1};Q_\tau)-\theta(x^{-1};Q_\tau)p,
  \\
  &{H'}^1_{\text{period}}(x,p)=\theta(p^{-1};Q_\tau)-x\theta(Q_m^{-1}p^{-1};Q_\tau).  
 \end{align}
\end{subequations}
The classical polynomials are consistent with the Seiberg--Witten curves of 6d U(1) gauge theory in~\cite{Hollowood:2003cv}. From these expressions, we find
\begin{subequations}
\begin{align}
&\mathsf{H}^1_{\text{period}}(q^{-1/2} Q_m^{-1} \mathsf{p},q^{1/2}\mathsf{x})\Big|_{Q_m\to q^{-1} Q_m^{-1}}={\mathsf{H}'}^1_{\text{period}}(\mathsf{x}, \mathsf{p}),
\\
&H^1_{\text{period}}(Q_m^{-1} p,x)\Big|_{Q_m \to Q_m^{-1}}={H'}^1_{\text{period}}(x,p).
\end{align}
\end{subequations}
 This means that the symplectic transformation exchanging $x$ and $p$ corresponds to the S-duality for the D5 and NS5 branes as shown in Fig.~\ref{period_chain}.
 Since $(x, p)$ exchange is realized as Fourier transformation, the corresponding open string partition functions, $\mathcal{Z}^{1,\text{o}}_\text{period}$ and $\mathcal{Z}^{'1,\text{o}}_\text{period}$, are expected to be converted to each other through Fourier transformation.

\begin{figure}[htb]
\centering
\includegraphics[width=10cm]{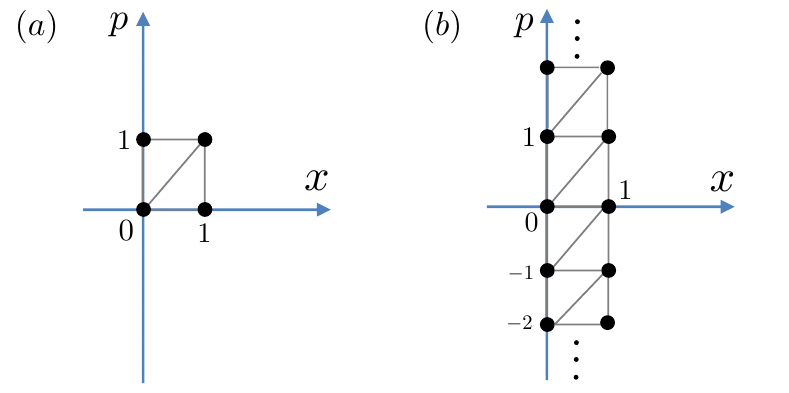}
\caption{The toric diagrams of the mirror curves. The left corresponds to the $1$-chain geometry, and the right panel corresponds to the periodic $1$-chain geometry.}
\label{toric1}
\end{figure}

From these expressions, the toric diagrams are given in Fig.~\ref{toric1}. Especially, we find that the toric diagram of the periodic chain geometry is given by the collection of the black dots infinitely spanned on one of the axis. Here we express the toric diagram of the periodic chain geometry as the dots spanned along the vertical axis.

\subsection{General case $N > 1$}\label{sec:N>1}
In the previous Section, we have provided the prescription to obtain the mirror curve, and demonstrated the computation in the simplest case. In this Section, we provide the result of the mirror curve in the general case. 

\begin{figure}[htb]
\centering
\includegraphics[width=10cm]{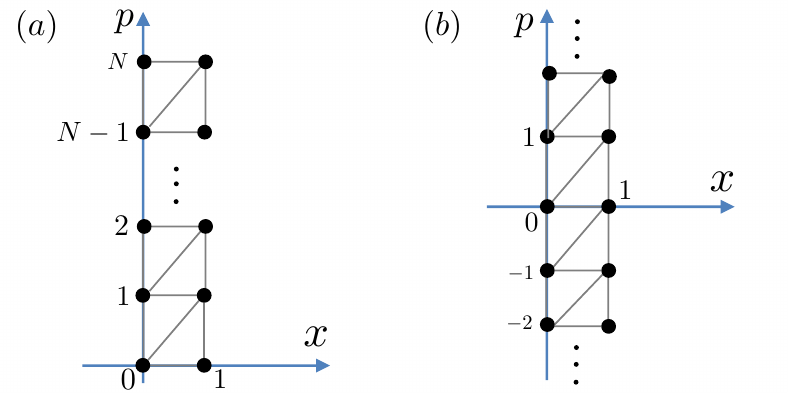}
\caption{The toric diagrams of the mirror curves in general case. The left panel corresponds to the $N$-chain geometry, and the right panel corresponds to the periodic $N$-chain geometry.}
\label{toric2}
\end{figure}

The partition function of the (periodic) $N$-chain geometries with single brane are given by
\begin{subequations}
\begin{align}
\mathcal{Z}_{\text{chain}}^{N,\text{o}}
&=
\prod_{a=1}^N 
\frac{(q^{1/2} x \prod_{k=1}^{a-1}Q_{\tau_k} ;q)_\infty}{( q^{1/2} x Q_{m_a}\prod_{k=1}^{a-1}Q_{\tau_k} ;q)_\infty},
\\
\mathcal{Z'}_{\text{chain}}^{N,\text{o}}
&=
\sum_{n=0}^\infty
(q^{1/2}x)^n
\prod_{a=1}^N
\frac{(Q_{m_a}\prod_{k=1}^{a-1} Q_{\tau_k}; q)_n }{(q\prod_{k=1}^{a-1} Q_{\tau_k}; q)_n},
\\
\mathcal{Z}_{\text{period}}^{N,\text{o}}
&=
\prod_{a=1}^N 
\frac{\Gamma_e (q^{1/2}x Q_{m_a} \prod_{k=1}^{a-1} Q_{\tau_k} ;q ;Q_\tau)}{\Gamma_e ( q^{1/2}x \prod_{k=1}^{a-1} Q_{\tau_k} ;q;Q_\tau)},
\\
\mathcal{Z'}_{\text{period}}^{N,\text{o}}
&=
\sum_{n=0}^\infty
(q^{1/2}x)^n
\prod_{a=1}^N
\frac{\theta(Q_{m_a}^{-1}\prod_{k=1}^{a-1} Q^{-1}_{\tau_k};q^{-1}; Q_\tau)_n }{\theta(q^{-1}\prod_{k=1}^{a-1} Q^{-1}_{\tau_k};q^{-1}; Q_\tau)_n}
.
\end{align}
\end{subequations}
We can easily find the quantum mirror curves by doing the same computation in Sec.~\ref{sec:N=1},
\begin{subequations}
\begin{align}
&\biggl[\prod_{a=1}^N (1-q^{1/2} \prod_{k=1}^{a-1}Q_{\tau_k}  \mathsf{x})\mathsf{p}-\prod_{a=1}^N (1- q^{1/2} Q_{m_a} \prod_{k=1}^{a-1}Q_{\tau_k} \mathsf{x})  \biggr]\mathcal{Z}_{\text{chain}}^{N,\text{o}}(x)=0,
\\
&\biggl[q^{1/2}\mathsf{x}\prod_{a=1}^N (1-Q_{m_a} \prod_{k=1}^{a-1}Q_{\tau_k}  \mathsf{p}) -\prod_{a=1}^N (1-\prod_{k=1}^{a-1}Q_{\tau_k}  \mathsf{p}) \biggr]{\mathcal{Z}'}_{\text{chain}}^{N,\text{o}}(x)=0,
\\
&\biggl[\prod_{a=1}^N \theta(q^{-1/2}\prod_{k=1}^{a-1} Q_{\tau_k}^{-1} \mathsf{x}^{-1};Q_\tau))\mathsf{p} - \prod_{a=1}^N \theta(q^{-1/2} Q_{m_a}^{-1}  \prod_{k=1}^{a-1} Q_{\tau_k}^{-1} \mathsf{x}^{-1} ;Q_\tau)\biggr]\mathcal{Z}_{\text{period}}^{N,\text{o}}(x)=0,
\\
&\biggl[q^{1/2}\mathsf{x}\prod_{a=1}^{N} \theta(\prod_{k=1}^{a-1} Q_{\tau_k}^{-1} Q_{m_a}^{-1} \mathsf{p}^{-1} ;Q_\tau) -\prod_{a=1}^N \theta(\prod_{k=1}^{a-1} Q_{\tau_k}^{-1}   \mathsf{p}^{-1};Q_\tau)) \biggr]{\mathcal{Z}'}_{\text{period}}^{N,\text{o}}(x)=0.
\end{align}
\end{subequations}
The quantum operators are 
\begin{subequations}
\begin{align}
&\mathsf{H}^N_{\text{chain}}(\mathsf{x},\mathsf{p})
=\prod_{a=1}^N (1- q^{1/2} \prod_{k=1}^{a-1}Q_{\tau_k} \mathsf{x})\mathsf{p}-\prod_{a=1}^N (1-q^{1/2} Q_{m_a} \prod_{k=1}^{a-1}Q_{\tau_k} \mathsf{x}),
\\
&{\mathsf{H}'}^N_{\text{chain}}(\mathsf{x},\mathsf{p})
=q^{1/2}\mathsf{x}\prod_{a=1}^N (1-Q_{m_a} \prod_{k=1}^{a-1}Q_{\tau_k}  \mathsf{p}) -\prod_{a=1}^N (1-\prod_{k=1}^{a-1}Q_{\tau_k}  \mathsf{p}),
\\
&\mathsf{H}^N_{\text{period}}(\mathsf{x},\mathsf{p})
=\prod_{a=1}^N \theta(q^{-1/2} \prod_{k=1}^{a-1} Q_{\tau_k}^{-1} \mathsf{x}^{-1};Q_\tau))\mathsf{p} - \prod_{a=1}^N \theta(q^{-1/2}Q_{m_a}^{-1} \prod_{k=1}^{a-1} Q_{\tau_k}^{-1}  \mathsf{x}^{-1} ;Q_\tau),
\\
&{\mathsf{H}'}^N_{\text{period}}(\mathsf{x},\mathsf{p})
=q^{1/2}\mathsf{x}\prod_{a=1}^{N} \theta(Q_{m_a}^{-1} \prod_{k=1}^{a-1} Q_{\tau_k}^{-1}  \mathsf{p}^{-1} ;Q_\tau) -\prod_{a=1}^N \theta(\prod_{k=1}^{a-1} Q_{\tau_k}^{-1}   \mathsf{p}^{-1};Q_\tau)),
\end{align}
\end{subequations}
and the classical polynomials are
\begin{subequations}\label{N-mirror_curves}
\begin{align}
&H_{\text{chain}}^N(x,p)
=p\prod_{a=1}^N (1-\prod_{k=1}^{a-1}Q_{\tau_k} x)-\prod_{a=1}^N (1-Q_{m_a} \prod_{k=1}^{a-1}Q_{\tau_k} x),
\\
&{H'}_{\text{chain}}^N(x,p)
=x\prod_{a=1}^N (1-Q_{m_a} \prod_{k=1}^{a-1}Q_{\tau_k} p) -\prod_{a=1}^N (1-\prod_{k=1}^{a-1}Q_{\tau_k} p),
\\
&H_{\text{period}}^N(x,p)
=p\prod_{a=1}^N \theta(\prod_{k=1}^{a-1} Q_{\tau_k}^{-1}   x^{-1};Q_\tau))-\prod_{a=1}^N \theta(Q_{m_a}^{-1} \prod_{k=1}^{a-1} Q_{\tau_k}^{-1} x^{-1} ;Q_\tau) ,
\\
&{H'}_{\text{period}}^N(x,p)
=x\prod_{a=1}^{N} \theta(Q_{m_a}^{-1} \prod_{k=1}^{a-1} Q_{\tau_k}^{-1} p^{-1} ;Q_\tau) -\prod_{a=1}^N \theta(\prod_{k=1}^{a-1} Q_{\tau_k}^{-1}  p^{-1};Q_\tau)).
\end{align}
\end{subequations}
When the K\"ahler parameters $Q_{m_a}$ are set to the same values, $Q_{m_a}=Q_m$, again the quantum operators and the classical polynomials have following relations,
\begin{subequations}
\begin{align}
&\mathsf{H}^N_{\text{chain}}(q^{-1/2}Q_m^{-1}\mathsf{p},q^{1/2}\mathsf{x})\Big|_{Q_m \to q^{-1} Q_m^{-1}}
={\mathsf{H}'}^N_{\text{chain}}(\mathsf{x},\mathsf{p}),\quad
H^N_{\text{chain}}(Q_m^{-1} p,x)_{Q_m \to  Q_m^{-1}}={H'}^N_{\text{chain}}(x,p),
\\
&\mathsf{H}^N_{\text{period}}(q^{-1/2}Q_m^{-1}\mathsf{p},q^{1/2}\mathsf{x})\Big|_{Q_m \to q^{-1} Q_m^{-1}}
={\mathsf{H}'}^N_{\text{period}}(\mathsf{x},\mathsf{p}),\quad
H^N_{\text{period}}(Q_m^{-1} p,x)_{Q_m \to  Q_m^{-1}}={H'}^N_{\text{period}}(x,p).
\end{align}
\end{subequations}
where we fix $Q_{\tau_k}$ under the shift $Q_m \to q^{-1} Q_m^{-1}$.
The toric diagrams of these mirror curves are given in Fig.~\ref{toric2}. In the $N$-chain geometry, the toric diagram of the mirror curve is extended to the collecton of $(2N+2)$ dots. These quantum mirror curves must be identified with the difference equations for the (elliptic) hypergeometric functions.

\begin{figure}[htb]
\centering
\includegraphics[width=14cm]{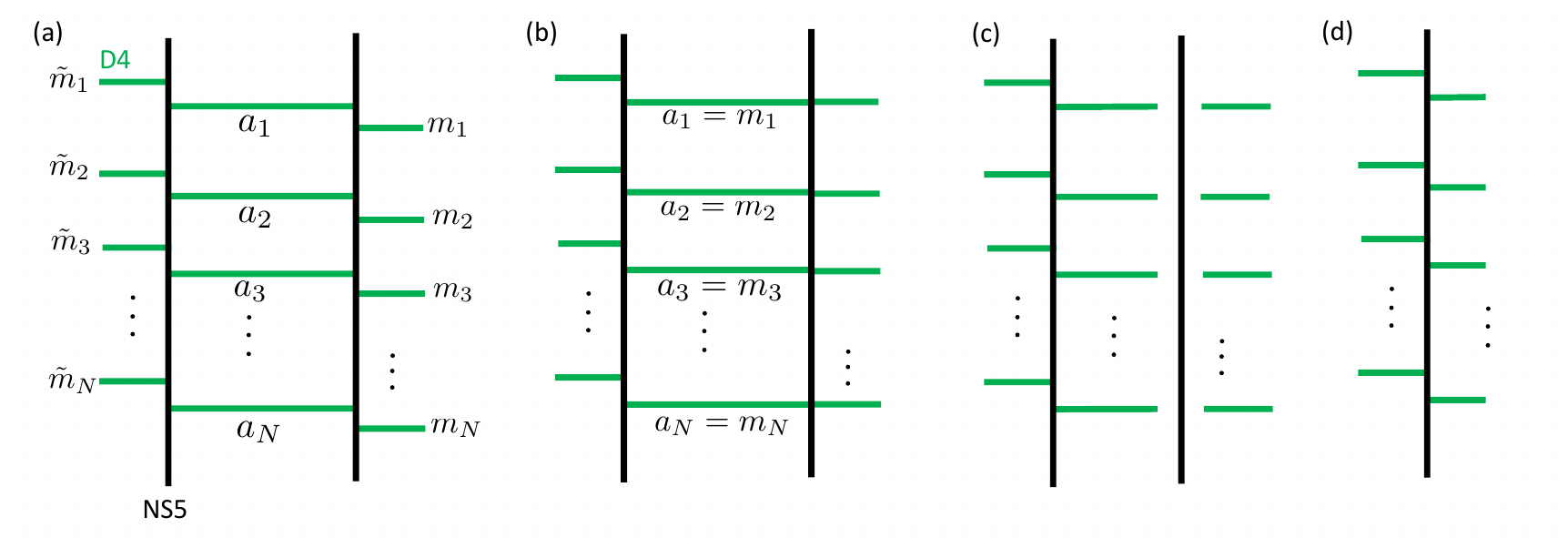}
\caption{Higgsing for the Hanany--Witten brane set up. By setting the Coulomb moduli parameters to the masses of the fundamental multiplets, we can remove one of the NS5-brane.}
\label{iia}
\end{figure}

 Let us connemt on a possible connection between the mirror curves \eqref{N-mirror_curves} and the Seiberg--Witten geometry.
 The Seiberg--Witten curve for 4d SU($N$) gauge theory with $N$ fundamental and anti-fundamental hypermultiplets, whose brane realization is found in Fig.~\ref{iia}, is given by
 \begin{align}
  \Sigma_\text{SW} = \{(x, p) \in \mathbb{C} \times \mathbb{C}^* \mid H_\text{SW}(x,p) = 0\}
 \end{align}
 with
 \begin{align}
  H_\text{SW}(x,p) = a(x) \, p + \frac{d(x)}{p} - T(x)
  \label{eq:SW_curve}
 \end{align}
 where $a(x)$, $d(x)$ are polynomials in $x$ of degree $N$, encoding the fundamental and anti-fundamental masses, $\{m_f, \tilde{m}_f\}_{f=1,\ldots,N}$,
 \begin{align}
  a(x) = \prod_{f=1}^N (x - m_f)
  \, , \qquad
  d(x) = \prod_{f=1}^N (x - \tilde{m}_f)
  \, .
 \end{align}
 The polynomial $T(x)$ encodes the Coulomb moduli $\{a_\alpha\}_{\alpha=1,\ldots,N}$.
 Equating the fundamental masses and the Coulomb moduli $a_\alpha = m_\alpha$ for $\alpha = 1, \ldots, N$, as known as the root of Higgs branch \cite{Dorey:1999zk}, the theory enters the Higgs branch and the right NS5 brane can be removed away.
 In the viewpoint of the Seiberg--Witten geometry, it corresponds to the factorization of the curve at $T(x) = a(x) + d(x)$,
 \begin{align}
  H_\text{SW}(x,p) = 0
  \quad \iff \quad
  \left( a(x) \, p - d(x) \right) \left( p - 1 \right) = 0
  \, .
 \end{align}
 Then we identify the first factor with the contribution of the left NS5 brane,
 \begin{align}
  a(x) \, p - d(x) = 0
  \, .
  \label{eq:SW_factor}
 \end{align}
 Turning on the $\Omega$-background, one can obtain the differential equation for the corresponding gauge theory partition function, which is regarded as quantization of \eqref{eq:SW_factor}~\cite{Fujimori:2015zaa}.
 
 We can also apply this argument to 5d and 6d theories.
 For 5d theory, we just replace $x \in \mathbb{C}$ with $\mathbb{C}^*$, and the factorized curve \eqref{eq:SW_factor} agrees with the mirror curve for the chain geometry \eqref{N-mirror_curves}.
 Actually the geometry shown in Fig.~\ref{iia} (d) coincides with the web diagram of the chain geometry. 
 For 6d theory, the polynomials, $a(x)$ and $d(x)$, are replaced with the elliptic function, and \eqref{eq:SW_factor} agrees with the periodic chain geometry similarly.

\subsection{Relation of the chain geometry and periodic chain geometry}
At a glance, the toric diagram of the $N$-chain geometry is similar with the one of the periodic $1$-chain geometry when $N$ is sufficiently large. Indeed, the partition function of the $\infty$-chain geometry with single brane under the constraints $Q_{\tau_k}=Q_\tau$ and $Q_{m_a}=Q_m$ is given by
\be
\ba
\mathcal{Z}_{\text{chain}}^{\infty,\text{o}}
&=
\prod_{n=1}^\infty \frac{(q^{1/2} x Q_\tau^{n-1} ;q)_\infty}{( q^{1/2} x Q_{m}Q_{\tau}^{n-1} ;q)_\infty},
\ea
\ee
and we find
\be
\ba
\mathcal{Z}_{\text{chain}}^{\infty,\text{o}}(x)\mathcal{Z}_{\text{chain}}^{\infty,\text{o}}(Q_\tau x^{-1})
=\mathcal{Z}_{\text{period}}^{1,\text{o}}(x).
\label{eq:o-p}
\ea
\ee
We interpret $\mathcal{Z}_{\text{chain}}^{\infty,\text{o}}(Q_\tau x^{-1})$ as the partition function of the topological strings with the anti-brane since the orientations of the worldsheets coming from left side and right side of the Lagrange submanifold are opposite. In terms of the toric diagram, this leads to the collection of the dots extended along the negative vertical axis.
Actually this observation suggests a possible realization of 6d (elliptic) gauge theory using 5d super group theory as discussed later.

\begin{figure}[htb]
\centering
\includegraphics[width=12cm]{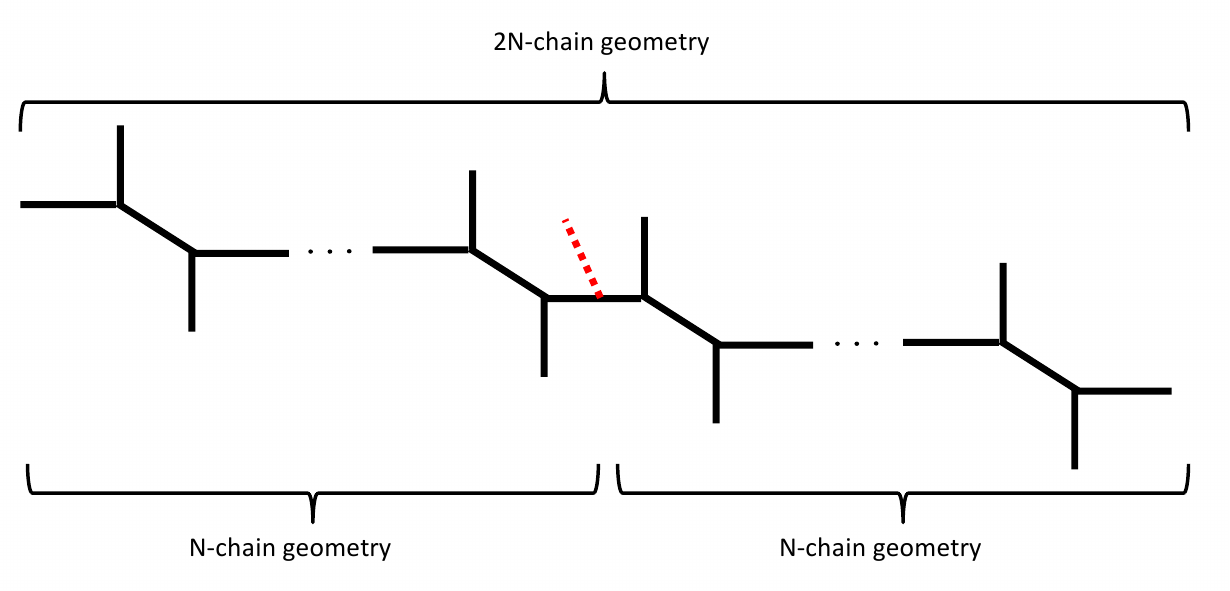}
\caption{The $2N$-chain geometry with the inner branes. The branes are located at the middle of the chain geometry.}
\label{chain_3}
\end{figure}

From this observation, when we calculate the partition function of the open topological string on the $2N$-chain geometry with the inner branes as in Fig.~\ref{chain_3}, and take the limit $N\to \infty$, we expect that this partition function agrees with the periodic $N$-chain geometry since the worldsheets attached on the brane from the left and right side have the opposite orientation, and correspond to the contributions coming from the brane and anti-brane.

The partition function is simply given by replacing the operator $(-Q_N)^{L_0}$ to $V_- (x_1) (-Q_N)^{L_0} V'_+ (x_2)$ in the partition function of the closed topological string \eqref{pf_chain_Nc},
\be
\ba
{\mathcal{Z}''}_{\text{chain}}^{N}
&=
\langle 0 | \prod_{a=1}^{N-1}  \biggl[ V_- (q^{-\rho})   V_+(q^{-\rho})    (-Q_{m_a})^{L_0}  
 V'_- (q^{-\rho}) V'_+(q^{-\rho})  (-Q_{a} )^{L_0}  \biggr] 
 \\
 &\qquad\times
 V_- (q^{-\rho})   V_+(q^{-\rho})    (-Q_{m_N})^{L_0}  
 V'_- (q^{-\rho}) V'_+(q^{-\rho})  V_- (x_1) (-Q_N)^{L_0} V'_+ (x_2)
 \\
 &\qquad\times
  \prod_{a=N+1}^{2N}  \biggl[ V_- (q^{-\rho})   V_+(q^{-\rho})    (-Q_{m_a})^{L_0}  
 V'_- (q^{-\rho}) V'_+(q^{-\rho})  (-Q_{a} )^{L_0}  \biggr] 
| 0 \rangle.
\ea
\ee
With the single brane, the partition function of the open topological strings is
\be
\ba
{\mathcal{Z}''}_{\text{chain}}^{N,\text{o}}
&=
\frac{{\mathcal{Z}''}_{\text{chain}}^{N}}{{\mathcal{Z}}_{\text{chain}}^{N,\text{c}}}
\\
&=
\prod_{a=1}^{N}\frac{(q^{1/2} x^{-1}Q_{m_a}^{-1}\prod_{k=a}^{N}Q_{\tau_k};q)_\infty}{(q^{1/2}x^{-1}\prod_{k=a}^{N}Q_{\tau_k};q)_\infty}
\times
\prod_{a=N}^{2N}\frac{(q^{1/2}x\prod_{k=N+1}^{a}Q_{\tau_k};q)_\infty}{(1-q^{1/2}xQ_{m_{a+1}}\prod_{k=N+1}^{a}Q_{\tau_k};q)_\infty}.
\ea
\ee
By setting $Q_{m_a}=Q_{m}$ and $Q_{\tau_a}=Q_\tau$, and taking $N\to\infty$, we find
\be
\ba
{\mathcal{Z}''}_{\text{chain}}^{N,\text{o}}
=
\prod_{n=1}^{\infty}
\frac
{(q^{1/2}xQ^{n-1}_{\tau_k};q)_\infty(q^{1/2}x^{-1}Q_{m}^{-1}Q^{n}_{\tau_k};q)_\infty}
{(q^{1/2} x^{-1} Q^n_{\tau} )_\infty(q^{1/2}xQ_{m} Q^{n-1}_{\tau};q)_\infty},
\ea
\ee
which is exactly the same as the partition function of the topological string with single brane on the periodic $1$-chain geometry, so that the mirror curves of two geometries are also the same.

Then, one might wonder what about the closed topological strings. The partition function on the $\infty$-chain geometry is given by setting $Q_{m_a}=Q_m$ and $Q_a=Q$, and taking $N\to \infty$ in \eqref{pf_chain_Nc2}  (as explained in, e.g.~\cite{Haghighat:2013tka}). The result is infinite product of the periodic $1$-chain geometry\footnote{One may read off the BPS spectrum from the asymptotic behavior of the partition function in the limit $N \to \infty$.},
\be
\ba
{\mathcal{Z}}_{\text{chain}}^{\infty,\text{c}}
=
\bigl({\mathcal{Z}}_{\text{period}}^{1,\text{c}}\bigr)^\infty.
\ea
\ee

The interpretation of the above results is as follows: When there is a $\mathbb{P}^1$'s, the worldsheet can be wrapped on 2-cycles whose K\"ahler parameter is $Q_{m}$, and the contribution is $(1-Q_{m}q^{i+j-1})$. Let us consider the case where there are some $\mathbb{P}^{1}$'s whose K\"ahler parameters are $Q_{m_a}$ and $Q_a$. Here we denote $\mathbb{P}^{1}$'s characterlized by $Q_{m_a}$ and $Q_a$ by $\mathbb{P}^{1}_{m_a}$'s and $\mathbb{P}^{1}_{a}$'s, respectively. In this situation, the worldsheet can be wrapped on some 2-cycles, and there are three types of contributions, $(1-Q_{m_a}\prod_{k=b}^{a-1} Q_{\tau_k} q^{i+j-1})$, $(1-Q_{m_b}\prod_{k=b+1}^a Q_{\tau_k} q^{i+j-1})$, and $(1-Q_{m_a}Q_b \prod_{k=b+1}^{a-1} Q_{\tau_k} q^{i+j-1})$, as in \eqref{pf_chain_Nc2}. When we fix $a=1$ in the product in \eqref{pf_chain_Nc2}, $b$ runs from 2 to $\infty$, and the collection of the product becomes ${\mathcal{Z}}_{\text{period}}^{1,c}$ under  $Q_{m_a}=Q_{m}$ and $Q_{\tau_a}=Q_\tau$, and these contributions come from the worldsheet wrapped on $\mathbb{P}^1_{m_a}$'s and $\mathbb{P}^{1}_{b}$'s where $a\geq1, b>1$. The similar discussion can be applied for $a\geq2$, so that the result is the infinite product of ${\mathcal{Z}}_{\text{period}}^{1,c}$.

In the open topological string, the partition function is devided by the partition function of the closed topological string, so that the contributions come from the worldsheet wrapped on $\mathbb{P}^{1}_{m_a}$'s and $\mathbb{P}^{1}_{a}$'s, and attached on the Lagrange submanifold. These contributions become to ${\mathcal{Z}}_{\text{period}}^{1,o}$. The important point is that the maps from the worldsheet to the toric Calabi--Yau threefold are connected, so that there is no other contribution due to the normalization, $\mathcal{Z}^{\text{o}}=\mathcal{Z}/\mathcal{Z}^{\text{c}}$.

The periodic $N$-chain geometry can be realized as the similar way: consider the $2NM$-chain geometry with single brane on the middle of the chain, and imposing the ``periodicity'' on the K\"ahler parameters as follows,
\be
\ba
Q_{m_{a+N}} =Q_{m_{a}},~Q_{\tau_{a+N}}=Q_{\tau_a}.
\ea
\ee
By taking the limit, $M\to\infty$, we obtain the partition function of the periodic $N$-chain geometry with single brane as the $\infty$-chain geometry with single brane.

The connection between the open and periodic chain geometry also implies a possible connection for the 6d and 5d gauge theories.
As shown in \eqref{eq:o-p}, we need to combine the brane and anti-brane contributions to realize the periodic chain geometry.
Since the anti-brane engineers the supergroup gauge theory~\cite{Vafa:2001qf}, this observation implies the connection between 5d supergroup gauge theory and 6d gauge theory.
We can see such a connection from the view point of Seiberg--Witten geometry.
The Seiberg--Witten curve for 6d $\mathcal{N} = (1,0)$ SU($N$) gauge theory on $\mathbb{R}^4 \times T^2$ is given by \eqref{eq:SW_curve} with the elliptic functions
\begin{align}
 T^\text{6d}_{\mathrm{SU}(N)}(x) = \prod_{\alpha=1}^N \theta(\re^{x - a_\alpha};p)
 \, , \qquad
 \text{etc}
\end{align}
where $p = \exp (2 \pi \ri \tau)$ is the elliptic nome with the torus modulus $\tau$.
Since the $\theta$-function is formally written as
\begin{align}
 \theta(z;p) = (z;p)_\infty (pz^{-1};p)_\infty = (z;p)_\infty(z^{-1};p^{-1})_\infty^{-1}
 \, ,
\end{align}
we have
\begin{align}
 T^\text{6d}_{\mathrm{SU}(N)}(x) =
 \prod_{\alpha=1}^N \prod_{n=0}^\infty \left( 1 - \re^{x - a_\alpha + 2 n \pi i \tau } \right) \left( 1 - \re^{- x + a_\alpha - 2 n \pi i \tau } \right)^{-1}
\end{align}
which coincides with the curve for 5d SU($NM|NM$) gauge theory with $M \to \infty$~\cite{Dijkgraaf:2016lym}.
It would be interesting to study such a Kaluza--Klein-like uplift of supergroup gauge theory from 5d to 6d for more details.

\section{Discussion}\label{discussion}
In this paper we have discussed the (quantum) mirror curves from the partition functions of the open topological strings calculated by the topological vertex. 

There are some topics for the further studies. First of all, it would be interesting to calculate the non-perturbative effects of the topological strings on the periodic chain geometry based on the mirror curve derived in this paper. According to \cite{Hatsuda:2013oxa}, the non-perturbative free energy of the topological strings on periodic $1$-chain geometry is given by the combination of the unrefined free energy and the Nekrasov--Shatashvili limit of the free energy of the  refined topological string.
As we mentioned in the introduction, the authors in \cite{Grassi:2014zfa} proposed that the non-perturbative free energy is closely related to the quantization of the mirror curve, and this conjecture is investigated in various examples. However, to the authors' knowledge, there is no evidence for the periodic geometry.

In this paper, we have obtained the genus-0 and -1 mirror curves corresponding to the chain geometry and the periodic chain geometry, respectively. The derivation of the higher genus mirror curve is difficult since there are two or more partitions that we cannot take the summation, so that we cannot derive the difference equation for the partition function at this stage. If this issue is resolved, we can investigate the B-model topological string theory on the more complicated chain geometry.

The mirror curve of the periodic chain geometry has been obtained from the infinitely long chain geometry. We can do the same process for the periodic chain geometry, namely, we can consider periodic $\infty$-chain geometry. By regarding $Q_\tau$ and $Q_{\tau_i}$ as a independent moduli parameters, and imposing $Q_{\tau_i}=p$, we obtain
\be
\ba
&\mathcal{Z}_{\text{period}}^{\infty,\text{o}} = \prod_{i=1}^\infty \frac{\Gamma_e (x q^{-i+1/2};Q_\tau;p)}{\Gamma_e (x Q_m  q^{-i+1/2};Q_\tau ; p)},
\ea
\ee
This result should be interpreted as ``compactification'' of the web diagram. 
It would be interesting to clarify what the physical meaning is.

In this paper, we have considered the topological strings with single brane. Due to the simplicity, the partition functions are expressed as the (elliptic) hypergeometric function. As a naive extension, we can consider the topological strings with multiple branes. Since the analysis is quite difficult due to the technical reason, it would be interesting to study this direction.

In the mathematical viewpoint, investigating the elliptic version of the $q$-binomial theorem is wothy. As we mentioned in Section \ref{mirror_curve}, the partition function $\mathcal{Z}_{\text{chain}}^{1,o}$ agrees with $1/{\mathcal{Z}'}_{\text{chain}}^{1,o}$ due to the symmetry of the geometry. When we compactify the geometry, the partition function is elliptized, and expressed as the theta function. Since the symmetry the chain geometry had is broken, some modification would be needed to find the relation between $\mathcal{Z}_{\text{period}}^{1,o}$ and ${\mathcal{Z}'}_{\text{period}}^{1,o}$.

We will come back to these issues in the near future.

\section*{Acknowledgement}

We would like to thank Andrea Brini for helpful discussion.
Y. S. was supported in part by the JSPS Research Fellowship for Young Scientists (No.~JP17J00828).
The work of TK was supported in part by Keio Gijuku Academic Development Funds, JSPS Grant-in-Aid for Scientific Research (No.~JP17K18090), MEXT-Supported Program for the Strategic Research Foundation at Private Universities ``Topological Science'' (No.~S1511006), JSPS Grant-in-Aid for Scientific Research on Innovative Areas ``Topological Materials Science'' (No.~JP15H05855), and ``Discrete Geometric Analysis for Materials Design'' (No.~JP17H06462).

\appendix
\section{Definitions and notations}\label{appA}
\subsection{Mathematical preliminaries}
Here we define several symbols.
\par
The $q$-shifted factorial ($q$-Pochhammer symbol) is defined as the product,
\be
\ba
(x;q)_n =
\begin{cases}
 1  \quad
   &\text{for} \quad n=0, \\ 
  \vspace{-5mm} \\ \displaystyle
 \prod_{k=0}^{n-1}(1-xq^k)  \quad
   &\text{for} \quad n\geq1, \\ 
   \vspace{-5mm}  \\ \displaystyle
   \prod_{k=0}^{-n}(1-xq^{-k})^{-1} \quad
   &\text{for} \quad n\leq-1. 
\end{cases}
\ea
\ee
The theta function is defined as the infinite products,
\begin{align}
 \theta(z;p) = (z;p)_\infty (pz^{-1};p)_\infty .
\end{align}
The elliptic gamma function is defined by
\be
\ba
\Gamma_e(x;p_1;p_2)
:=\prod_{m,n\geq0}\frac{1-x^{-1}p_1^{n+1}p_2^{m+1}}{1-x p_1^{n}p_2^{m}}
={\rm exp}\Biggl[ \sum_{m\neq 0}\frac{x^m}{m(1-p^m)(1-q^m)}\Biggr],
\ea
\ee
which relates to the theta function,
\begin{subequations}
\begin{align}
&\Gamma_e (p_1 x) = \theta(x;p_2)\Gamma_e (x),
\\
&\Gamma_e (p_2 x) = \theta(x;p_1)\Gamma_e (x).
\end{align}
\end{subequations}
The hypergeometric function $_r \phi_{s-1}(a_1,a_2,...,a_r; b_1,b_2,...,b_s ; q; z)$ and its elliptization, the elliptic hypergeometric function $_r E_{r-1} (a_1,a_2,...,a_n;b_1,b_2,...,b_{n-1};z)$, are defined as
\begin{subequations}
\begin{align}
& _r \phi_{s-1}(a_1,a_2,...,a_r; b_1,b_2,...,b_s ; q; z)
\nonumber \\
&\qquad =
 \sum_{n=0}^\infty
  \frac{z^n}{(q;q)_n}\bigl( (-1)^n q^{1/2 n(n-1)} \bigr)^{1+s-r}
 \frac{(a_1;q)_n (a_2;q)_n \cdots (a_r;q)_n}{(b_1;q)_n (b_2;q)_n \cdots (b_s;q)_n},
 \\
& _r E_{r-1} (a_1,a_2,...,a_n;b_1,b_2,...,b_{n-1};z)
\nonumber \\ 
&\qquad =
\sum_{n=0}^\infty
z^n
\frac{\theta(a_1 ;q ;p)_n \theta(a_2 ;q ;p)_n \cdots \theta(a_r ;q ;p)_n}
{\theta(q ;q ;p)_n \theta(b_1 ;q ;p)_n \theta(b_2 ;q ;p)_n \cdots \theta(b_{r-1} ;q ;p)_n}.
\end{align}
\end{subequations}

\subsection{Topological Vertex}
\begin{figure}[htb]
\centering
\includegraphics[width=4cm]{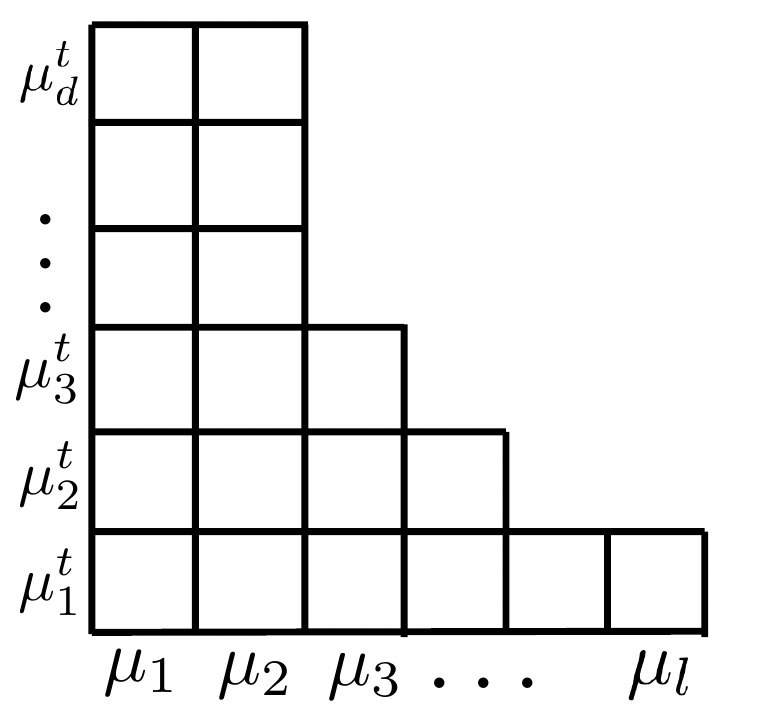}
\vspace{-5mm}
\caption{The Young diagram we use in this paper.}
\label{young}
\end{figure}

To define the topological vertex, first we define the Young diagram as the collection of boxes in Fig.~\ref{young}, which is a graphical representation of the partition.
We denote the variables $\mu_i$ and $\mu_j^t$ as the number of boxes $i$-th column and $j$-th row. In the Young diagram we define the coordinate $(i,j)\in\mu$ as $i=\mu_i,~j=\mu^t_j$. We also define some quantities associated with the Young diagram as follows,
\be
\ba
|\mu|=\sum_{i=1}^l \mu_i,~||\mu||^2 = \sum_{i=1}^l \mu_i^2.
\ea
\ee

Now we ready to define the topological vertex which is labelled by the Young diagram,
\be
\ba
&C_{\lambda \mu \nu}(q) = q^{\frac{\kappa_{\mu}+\kappa_{\nu}}{2}}s_{\nu}(q^{-\rho})
\sum_{\eta} s_{\lambda^{t}/\eta}(q^{-\nu-\rho})s_{\mu/\eta}(q^{-\rho-\nu^{t}}),
\\
&\qquad\kappa_\mu = ||\mu||^2 -||\mu^t||^2,~\rho = -1/2,-3/2,-5/2,\cdots,
\ea
\ee
where $s_\mu(x)$ is symmetric function called as Schur function, and $s_{\mu/\eta}(x)$ is the skew-Schur function defined as
\be
\ba
s_{\mu/\eta}(x) = \sum_{\lambda}N^\mu_{\eta\lambda}s_{\lambda}(x),
\ea
\ee
where $N^\mu_{\eta\lambda}$ is the Littlewood--Richardson coefficient.

\subsection{Vertex operators}
\begin{figure}[htb]
\centering
\includegraphics[width=4cm]{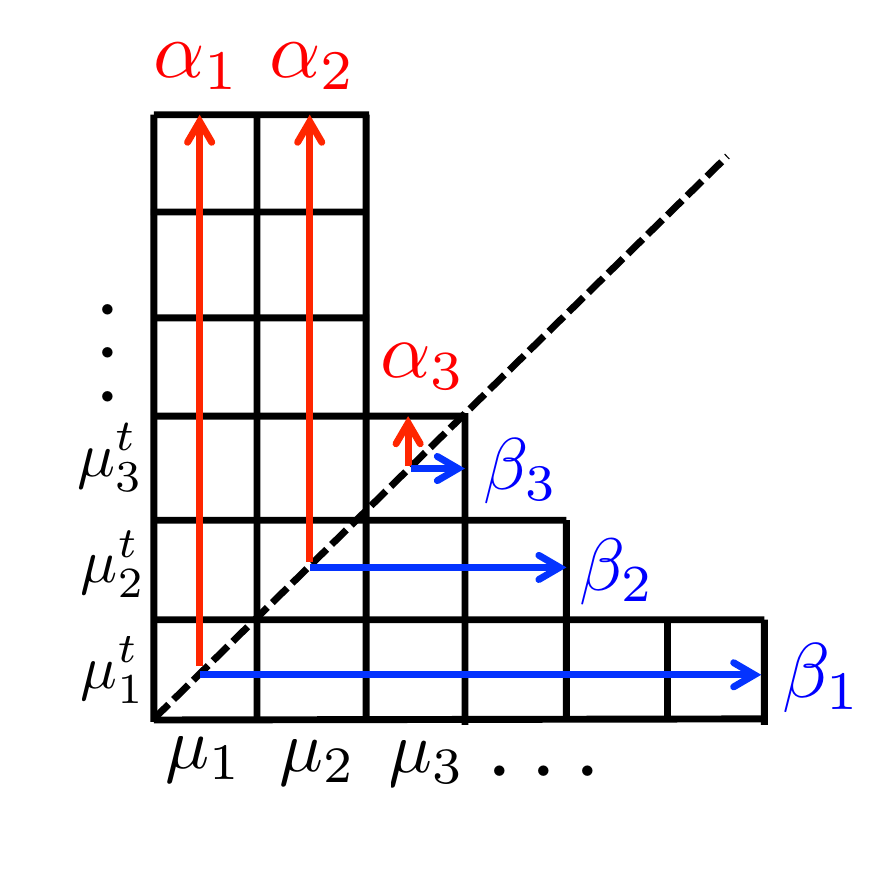}
\vspace{-5mm}
\caption{The Frobenius coordinate. By drawing the diagonal line, we can define two kinds of the lengths, $\alpha_i$ and $\beta_i$.}
\label{frob}
\end{figure}

The Schur function can be expressed using the vertex operators,
\be
\ba
&s_{\mu/\nu}(x)=\langle\nu |V_+(x)|\mu\rangle = \langle\mu |V_-(x)|\nu\rangle,
~
&s_{\mu^t/\nu^t}(x)=\langle\nu |V'_+(x)|\mu\rangle = \langle\mu |V'_-(x)|\nu\rangle,
\ea
\ee
where the operators $V_\pm(x),~V'_\pm(x)$ are the vertex operators with the bosonic operators,
\be
\ba
&V_{\pm}(x)={\rm exp}\biggl[ \sum_{n=1}^{\infty}\frac{J_{\pm n}}{n}\sum_{i=1}^{\infty}x_i ^n\biggr]
=\prod_{i=1}^{\infty}V_{\pm}(x_i),
~
&V'_{\pm}(x)={\rm exp}\biggl[ -\sum_{n=1}^{\infty}\frac{J_{\pm n}}{n}\sum_{i=1}^{\infty}(-x_i) ^n\biggr]
=\prod_{i=1}^{\infty}V'_{\pm}(x_i),
\\
& [J_n, J_m]=n\delta_{n+m,0}.
\ea
\ee
We remark $V'_\pm(x) = V_{\pm}(-x)^{-1}$.
The state $|\mu\rangle$ is defined as the product of the fermionic operators,
\be
\ba
|\mu\rangle=\prod_{i=1}^d\psi_{-\alpha_i}\prod_{i=1}^d\psi_{-\beta_i}^*|0\rangle, \quad
\psi_n|0\rangle=\psi_n^*|0\rangle=0
\quad \text{for} \quad n>0,
\ea
\ee
with 
\begin{align}
 \left\{ \psi_n, \psi^*_m \right\} = \delta_{n+m,0}
\end{align}
where $\alpha_i$ and $\beta_i$ are the lengths defined in the Frobenius coordinate (see Fig.~\ref{frob}), and satisfies the completeness relation,
\be
\ba
\sum_\mu |\mu \rangle \langle \mu | = {\bm 1}.
\ea
\ee
The number of the boxes in the Young diagram $|\mu|$ is counted by the operator
\be
\ba
|\mu||\mu\rangle = L_0|\mu\rangle ,~
L_0 = \frac{1}{2}J_0 ^2 +\sum_{n=1}^{\infty}J_{-n}J_n.
\ea
\ee

The commutation relations for $V_{\pm}(x)$, $V'_{\pm}(x)$, and $L_0$ are given as following,
\begin{subequations}
\begin{align}
&V_+(x)V'_-(y)=
\prod_{i,j=1}^{\infty}(1+x_i y_j)
V'_{-}(y)V_{+}(x),
\\
&V'_{+}(x)V_{-}(y)=
\prod_{i,j=1}^{\infty}(1+x_i y_j)
V'_{-}(y)V_{+}(x),
\\
&V_{+}(x)V_{-}(y)=
\prod_{i,j=1}^{\infty}(1-x_i y_j)^{-1}
V_{-}(y)V_{+}(x),
\\
&V'_{+}(x)V'_{-}(y)=
\prod_{i,j=1}^{\infty}(1-x_i y_j)^{-1}
V_{-}(y)V_{+}(x),
\\
&q^{L_0}V_{\pm}(x)q^{-L_0}=V_{\pm}(q^{\mp 1}x),
\\
&q^{L_0}V'_{\pm}(x)q^{-L_0}=V'_{\pm}(q^{\mp 1}x).
\label{VO_comm}
\end{align}
\end{subequations}
They are useful to calculate the partition functions of the topological string.





\providecommand{\href}[2]{#2}\begingroup\raggedright\endgroup

\end{document}